# Localized Exciton Emission with Spontaneous Circular Polarization in NiPS$_3$/WSe$_2$ Heterostructures

Adi Harchol[1], Shahar Zuri[1], Rajesh Kumar Yadav[2], Nirman Chakraborty[1,3], Idan Cohen[1], Tomasz Woźniak[4], Thomas Brumme[5], Thomas Heine[5,6,7*], Doron Naveh[2*] and Efrat Lifshitz[1*]

1. Schulich Faculty of Chemistry, Solid State Institute, Russell Berrie Nanotechnology Institute, Helen Diller Quantum Information Center, Technion, Haifa 3200003, Israel
2. Faculty of Engineering, Institute for Nanotechnology and Advanced Materials, Bar-Ilan University, Ramat-Gan 52900, Israel.
3. Eduard Zintl Institute of Inorganic and Physical Chemistry, Technical University of Darmstadt, 64287 Darmstadt, Germany
4. University of Warsaw, Faculty of Physics, 02-093 Warsaw, Poland
5. Faculty of Chemistry and Food Chemistry, Technische Universität Dresden, 01062 Dresden, Germany.
6. Helmholtz-Zentrum Dresden-Rossendorf, Center for Advanced Systems Understanding, CASUS, Untermarkt 20, 02826 Görlitz, Germany.
7. Department of Chemistry and ibs for Nanomedicine, Yonsei University, Seodaemun-gu, Seoul, 120-749 South Korea.

\* Corresponding authors: ssefrat@technion.ac.il, doron.naveh@gmail.com, thomas.heine@tu-dresden.de

**Abstract**

Two-dimensional (2D) van der Waals (vdW) heterostructures (HSs) provide a versatile platform for tailoring electronic, optical, and magnetic properties via proximity effects at their interfaces. In this work, we explore the optical response of few-layer NiPS$_3$/WSe$_2$ HSs using low-temperature micro-photoluminescence (μ-PL) and magneto-PL spectroscopy. The HSs exhibit multiple sharp excitonic peaks that do not appear in the individual constituent materials, indicating the emergence of localized intralayer WSe$_2$ excitons confined by interface-induced potentials. Notably, these excitons exhibit spontaneous circular polarization even in the absence of an external magnetic field, suggesting a magnetic proximity effect induced by uncompensated spins at the NiPS$_3$ interface. Magneto-PL measurements further reveal nonlinear Zeeman splitting, consistent with the presence of an interfacial exchange field that alters the valley exciton dynamics. Density functional theory (DFT) calculations confirm the intralayer origin of the PL and reveal interfacial hybridization and spin texture modifications, supporting the experimental findings. These results highlight how combining a 2D semiconductor with a layered antiferromagnet enables control over valley polarization and spin degrees of freedom, offering new opportunities for chiral light sources and magnetically tunable optoelectronic devices.

**Keywords:** NiPS$_3$/WSe$_2$ heterostructures, magnetic proximity effect, valley polarization, 2D materials, interface-induced potential.

## 1. Introduction

In the last two decades, two-dimensional (2D) van der Waals (vdW) materials have garnered substantial interest due to their potential for tailoring properties via the formation of heterostructures (HSs). The physical properties of the HSs constituent materials can be strongly influenced by proximity effects.[1–3] The transition metal dichalcogenides (TMDs) semiconductors, such as tungsten diselenide ($WSe_2$), have garnered immense interest for their direct bandgap in the monolayer limit, tightly bound excitons, and strong spin-valley coupling.[4–8] In each monolayer, tungsten (W) atoms are sandwiched between two selenium (Se) layers, with strong covalent bonding in-plane and comparatively weak vdW interactions out-of-plane (Figure 1a). $WSe_2$ crystalizes in the hexagonal $P6_3/mmc$ space group, with lattice constants of $a = b = 0.33$ nm and $c = 1.37$ nm.[9] Monolayer $WSe_2$ features broken inversion symmetry, giving rise to two degenerate yet inequivalent valleys (K and K') at the

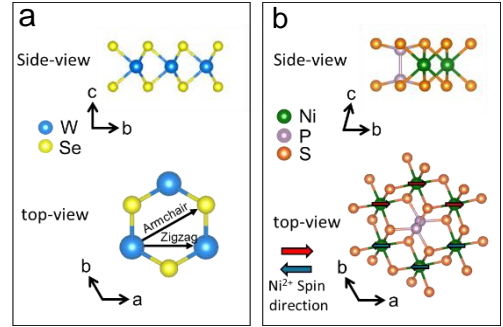

**Figure 1. a.** side-view (top) and top-view (bottom) of the $WSe_2$ and **b.** $NiPS_3$ crystal structure. Red and blue arrows represent two oppositely aligned spins of $Ni^{2+}$ ions, corresponding to the AFM zigzag configuration. The armchair and zigzag crystallographic directions of both materials are marked by black arrows in the top-view illustration of $WSe_2$.

Brillouin zone corners, which can be selectively excited using circularly polarized light.[10] This valley-selective selection rule underlies the field of valleytronics, in which the valley index of carriers serves as a new degree of freedom for information processing. Recent studies have leveraged these properties for valleytronics, demonstrating that applying a magnetic field (Zeeman effect) lifts the valley degeneracy and yields measurable valley polarization in photoluminescence (PL) spectroscopy.[11] However, the Zeeman-induced splitting in bare monolayer $WSe_2$ is relatively small (~0.2 meV/T),[12] motivating alternative routes for leveraging the valley polarization via a proximity to a magnetic vdW layer.[13–16]

An intriguing magnetic vdW compound is the nickel phosphorus trisulfide ($NiPS_3$), which belongs to the broader $MPX_3$ family (M = Fe, Ni, Mn; X = S, Se) of layered antiferromagnets. $NiPS_3$ adopts a monoclinic structure (space group C2/m), with lattice parameters $a = 0.58$ nm, $b = 1.01$ nm, $c = 0.66$ nm, and $\beta = 107°$.[17] Within each layer, Ni atoms form a near-honeycomb network coordinated by sulfur (S) and phosphorus (P) atoms (Figure 1b). Below its Néel temperature (~155 K), $NiPS_3$ exhibit zigzag in-plane antiferromagnetic (AFM) order, with a slight canting (~8°) out of plane (Figure 1b, bottom).[18,19] This magnetic arrangement gives rise to various electronic-magnetic phenomena,[20–22] including a linearly polarized PL band at 1.476 eV.[23–28]

Therefore, combining a semiconducting TMDs with a magnetic vdW crystal, whether ferro- or antiferromagnetic, enables a vast number of new phenomena,[1,2,29–31] due to strain, interfacial spin-orbit interactions and magnetic proximity.[3,32–37] The placing of a monolayer TMD on FM substrate such as EuS or $CrI_3$[14,38–41] showed that a proximity field with an effective magnetic field of tens of tesla, lifted the valley degeneracy by several meV.[13–16] Recent work on few-layer $WS_2$ on the AFM $FePS_3$ exhibited the emergence of an interfacial FM and sizeable valley splitting.[42] Furthermore, chiral quantum emitters have been reported in $WSe_2/NiPS_3$ under locally induced strain.[36] Yet, a comprehensive study of unstrained $NiPS_3/WSe_2$ HSs, particularly in the few-layer regime, remains relatively unexplored.

In this work, we report the optical and magneto-optical response of few-layer $NiPS_3/WSe_2$ HSs. Low-temperature micro photoluminescent (μ-PL) spectroscopy uncovers the emergence of multiple sharp peaks, absent in either parent crystal, which are attributed to localized excitons likely confined by an interface potential. Strikingly, the emission from these localized states exhibits strong circular polarization at zero external magnetic field, pointing to a valley polarization induced by the interfacial proximity of WSe$_2$ to uncompensated spins at the NiPS$_3$ interface. Density functional theory (DFT) calculations confirm the intralayer nature of the optical transition and reveal a spatially varying potential that acts as a trapping site for the exciton. In addition, the calculations show strong interfacial orbital hybridization between WSe$_2$ and NiPS$_3$, which modifies the spin texture at the interface. This study expands the horizon of vdW HSs by combining a 2D semiconductor with an antiferromagnet to enable controllable spin-valley effects and chiral excitonic emitters.

## 2. Results and Discussion

2.1. **Fabrication and Characterization.** Figure 2a presents optical microscope images of the investigated HSs, fabricated via a dry-transfer technique. *Device 1* (HS1, top panel in Figure 2a), has a stacked sequence of Si/SiO$_2$, hBN, NiPS$_3$, and WSe$_2$. *Device 2* (HS2, bottom panel in Figure 2a) incorporates Si/SiO$_2$, Au electrodes, NiPS$_3$, WSe$_2$, and a capping hBN layer. The Au electrodes in HS2 facilitate photoconductivity measurements, details of which will be published separately. In both devices, the NiPS$_3$ flakes are multilayer, while the WSe$_2$ flakes consist of a few layers - 35 layers in HS1 and 10 layers in HS2 - as determined by atomic force microscopy (AFM) (see supporting information (SI), Figure S1a,b). Figure 2b shows Raman spectra acquired at three distinct locations on *Device 2*: pure NiPS$_3$, pure WSe$_2$, and the overlapping NiPS$_3$/WSe$_2$ region. The Raman modes of both NiPS$_3$ and WSe$_2$ align well with previously reported values for unstrained crystals,[43,44] confirming the high quality of the exfoliated flakes. The spectrum of the HS region exhibits the characteristic Raman peaks of WSe$_2$ along with weak additional peaks (marked with asterisks) corresponding to the P$_2$, P$_5$, P$_7$ and P$_8$ modes of NiPS$_3$, confirming the presence of both materials in the stacked region. Similar Raman results for *Device 1* are shown in the SI, Figure S1c. Essentially, the main vibrational modes of WSe$_2$ in the HS region show no discernible shift, broadening or splitting compared to the exposed WSe$_2$, providing strong evidence for the absence of strain in the HS (see Section 1.1 in the SI for peak fitting and analysis). Figure 2c displays polarized second harmonic generation (P-SHG) plots from the WSe$_2$ (blue) and NiPS$_3$ (pink) regions in *Device 1* (top) and *Device 2* (bottom), together with theoretical fits (see

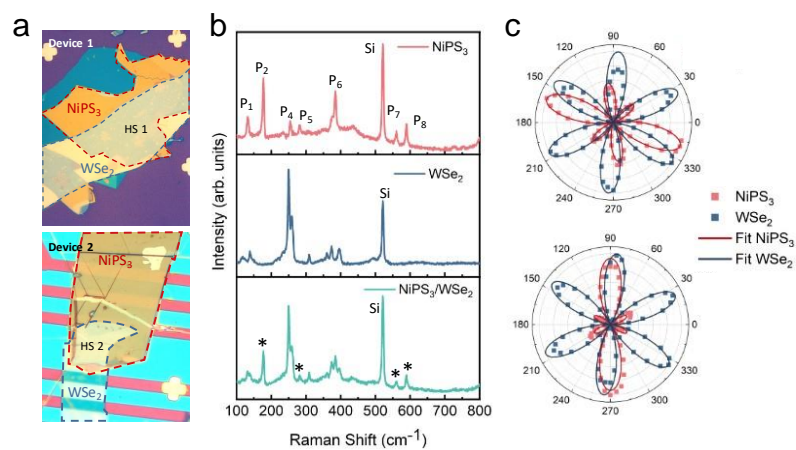

**Figure 2. a.** Optical microscope images of *Device 1* (top) and *Device 2* (bottom) **b.** Raman spectra of *Device 2* recorded at three different locations on top of the sample: NiPS$_3$, WSe$_2$, and NiPS$_3$/WSe$_2$ HS (from top to bottom), measured under 532 nm laser excitation at room temperature. **c.** P-SHG polar plots acquired from NiPS$_3$ (pink squares) and WSe$_2$ (blue squares) regions of *Device 1* (top) and *Device 2* (bottom), shown together with fitted curves (solid lines).

Section 1.2 in the SI for more details), showing good agreement with the expected angular dependence. Consistent with prior studies,[45,46] few-layer WSe$_2$ shows a six-fold symmetry, reflecting the threefold rotational symmetry. In contrast, a loss of six-fold symmetry is observed for the NiPS$_3$, presumably due to its monoclinic lattice and stacking offset between adjacent honeycomb layers.[45] WSe$_2$ and NiPS$_3$ possess distinct crystal structures, leading to strongly incommensurate HSs. As a result, the HSs are not expected to reconstruct into metastable stacking configurations, as commonly observed in commensurate or nearly commensurate TMD HSs and the formation of a well-defined moiré superlattice is therefore unlikely.

2.2. **PL Measurements.** Low temperature (4 K) μ-PL spectroscopy was employed to probe the optical emission across different regions of the sample. Measurements were performed under unpolarized excitation using a 480 nm continuous-wave laser and were collected at the center of each region, far from any flake edges, to exclude edge-related defect emission. Figure 3a compares the PL spectra of exposed NiPS$_3$ (red line) and the overlapped NiPS$_3$/WSe$_2$ region in both HSs, presented by the blue and green lines. The PL of the exposed NiPS$_3$ region depicts a sharp emission band at 1.476 eV, matching reported excitonic transitions in NiPS$_3$.[21,23,26,47] This emission exhibits strong linear polarization (inset to Figure 3c), consistent with NiPS$_3$'s in-plane anisotropy. Conversely, the bare few-layer WSe$_2$ region shows negligible PL, characteristic of forbidden optical transition in few-layer WSe$_2$ (Figure S3 in the SI).[48,49] This confirms that the strong emission in the HSs originates only from the NiPS$_3$/WSe$_2$ interface, and not from the individual WSe$_2$ or NiPS$_3$ flakes alone.

In contrast to either individual material,[21,48] the emission spectra of the HSs exhibit a series of discrete PL peaks spanning the range of 1.48 - 1.56 eV (blue and green curves in Figure 3a). Specifically, HS1 exhibits up to seven narrow emission lines, including a closely spaced group (p1-p3) with mutual energy separations of ~7 meV, and a second set of lines (p1, p4, p5) with energy separation of ~35 meV between consecutive peaks. A weak NiPS$_3$ emission line at 1.476 eV is also observed in the PL spectrum of HS1 (marked with an asterisk). In addition, a faint peak near 1.63 eV (also marked with an asterisk) appears, but is not discussed any further here. The PL spectrum of HS2 exhibits a primary emission line at ~1.56 eV (p6), along with satellite peaks located ~20 meV and ~40 meV below the main emission band. The sharp lines observed in the PL spectra of the HSs are attributed to localized excitons confined by interface-induced potentials or defect-induced traps at the NiPS$_3$/WSe$_2$ interface[36,50–53]. These features resemble the multi-peak emission patterns reported for moiré-trapped excitons in TMD bilayers or localized excitons at potential minima.[54,55] Further discussion of the multi-line origin is provided below. In addition, the intensity of these peaks significantly decreased above 50 K (see Figure S4 and Section 2.2 in the SI). In additional heterostructures of NiPS$_3$/WSe$_2$ (denoted as HS3 and HS4), not discussed in detail here, a sharp emission peak at 1.56 eV was also observed (Figure S8 in the SI), consistent with the emission features seen in both HS1 and HS2, which further confirms the reproducibility of the optical response in the NiPS$_3$/WSe$_2$ HSs.

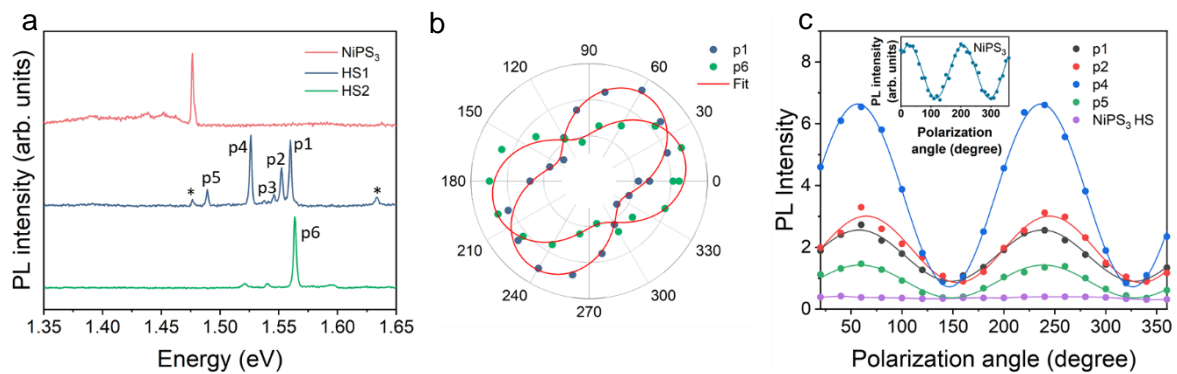

**Figure 3. a.** PL spectra of an exposed NiPS$_3$ region (pink), HS1 (blue), and HS2 (green), excited by 480 nm laser at 5 K. **b.** Polar plots of PL intensity under linear polarization detection for the p1 (blue) and p6 (green) peaks. The red solid lines represent fits to the data. **c.** Linear polarization dependence of PL intensity for peaks p1-p5 in HS1. Inset: Linear polarization dependence of an exposed NiPS$_3$ region. Solid lines represent fits to the data, except for p5, where the solid line serves only as a visual guide.

**Polarization-Resolved PL Measurements.** Figure 3b presents the linear polarization of the 1.56 eV emission peak of HS1 (p1) and HS2 (p6). The intensity versus polarization angle ($\theta$) follow the relation: $I_0 + I_d \cdot \sin^2(\theta - \theta_{max})$, where $I_0$ is the minimum PL intensity, $I_d$ is the difference between minimum and maximum intensity, and $\theta_{max}$ is the angle corresponding to the maximum intensity (solid lines in Figures 3b and 3c). It is important to note that both devices were not aligned along a specific crystallographic axis of either NiPS$_3$ or WSe$_2$; hence, the polarization angles of P1 and P6 (Figure 3b) are arbitrary. Additional peaks in the PL spectrum of HS1 also exhibit pronounced linear polarization, as shown in Figure 3c. This behavior is likely associated with inversion symmetry breaking at the interface, which induces a preferred excitonic dipole orientation. Moreover, the in-plane magnetic anisotropy of bulk NiPS$_3$ may contribute to stabilizing the exciton emission along specific crystallographic direction.

However, the NiPS$_3$ characteristic linear polarization of the 1.476 eV emission band found in pure NiPS$_3$ slabs (inset, Figure 3c) is absent in the related PL peak of the heterostructure (Figure 3c, purple circles). This suppression is likely due to the weak emission signal of NiPS$_3$ in the heterostructure, which is often difficult to resolve above the background noise, as shown by the polarization-resolved PL spectra in Figure S5 in the SI.

Notably, circularly polarized detection of the heterostructure emission bands at zero external magnetic field reveals an intrinsic helicity. Figure 4a depicts the PL spectra of HS1 (top) and HS2 (bottom), measured under circular polarization detection. The detection path was calibrated using the PL emission of bare NiPS$_3$, which exhibits strong linear polarization but negligible circular polarization (see Figure S6 in the SI). The absence of circular polarization in this reference confirms that the optical setup does not introduce artificial helicity. The optical alignment was further verified by the systematic evolution of the circular polarization with the applied magnetic field (shown below), confirming its intrinsic origin. Remarkably, even at zero applied magnetic field, the HSs show a preferred helicity in their emission, suggesting an effective interfacial exchange field arising from uncompensated or canted spins at the NiPS$_3$ interface. Although bulk NiPS$_3$ is an antiferromagnet with zigzag in-plane ordering, perfect magnetic compensation may not be preserved at surfaces or interfaces, where the absence of neighboring NiPS$_3$ layers and hybridization with WSe$_2$ can modify the spin configuration.[42] The opposite helicity observed in HS1 and HS2 likely reflects variations in the interfacial spin configuration of the NiPS$_3$ flakes between independently assembled heterostructures. To assess spatial uniformity, circularly polarized PL measurements were performed at multiple locations across the HSs (Figure S7 in the SI), all yielding the same helicity and consistent spectral features, with minor variations in peaks intensities. This result contrasts with prior studies of monolayer WSe$_2$/NiPS$_3$ HSs, which reported negligible circular polarization at B = 0 T in the unstrained regime.[36] The discrepancy likely originates from the WSe$_2$ thickness: in Ref.[36] the PL was dominated by broad emission from monolayer WSe$_2$, which can overlap with or mask weaker interfacial emission. In contrast, the present HSs incorporate few-layer WSe$_2$, which exhibits negligible PL at 5 K (Figure S3 in the SI), thereby enabling direct observation of emission arising from the NiPS$_3$/WSe$_2$ interface.

## 2.3. Magneto-PL Measurements.

Magneto-PL measurements were performed on the investigated HSs by applying an external out-of-plane magnetic field ($\vec{B} \perp \hat{a}, \hat{b}$) while measuring the emission under circularly polarized detection normal to the $\hat{a}, \hat{b}$ plane. Figures 4b and 4c depict the circularly resolved emission spectra for HS1 and HS2, respectively, recorded at different magnetic field strengths. The degree of circular polarization (DCP) is defined as:

(1) $\rho = \frac{|I_{\sigma^+} - I_{\sigma^-}|}{I_{\sigma^+} + I_{\sigma^-}}$

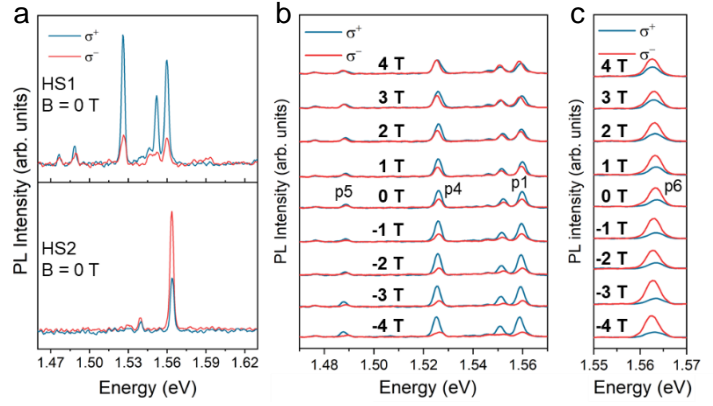

**Figure 4. a.** PL spectra of HS1 (top) and HS2 (bottom) recorded under circularly polarized detection at zero external magnetic field. **b, c.** Circularly resolved PL spectra of HS1 (**b**) and HS2 (**c**) measured under varying out-of-plane magnetic fields. The measurements were performed under 488 nm laser excitation.

where $I_{\sigma^+}$ ($I_{\sigma^-}$) is the PL intensity of the $\sigma^+$ ($\sigma^-$) component, is plotted vs. B for HS1 and HS2 in Figures 5a and 5b, respectively. In HS1 (HS2), negative fields yield DCP of ~0.7 (~0.6), whereas at positive fields the emission becomes nearly unpolarized (DCP ≈ 0).

Figures 5c and 5d extract the energy shift of the $\sigma^+$ and $\sigma^-$ components ($E_{\sigma^+}$ and $E_{\sigma^-}$), relative to the zero-field energy of the $\sigma^+$ component, for HS1's p1 and HS2's p6 peaks, respectively. The results reveal an asymmetric signature in the energy shift of the $\sigma^+$ and $\sigma^-$ branches in both HSs. Subtracting these energies yields a Zeeman split ($\Delta E = E_{\sigma^+} - E_{\sigma^-}$) as shown in Figures 5e and 5f. A purely linear Zeeman effect would produce a straight line against B with slope proportional to an effective g-factor ($\Delta E = g_{eff} \mu_B B$, where $\mu_B$ is the Bohr magneton), where a typical g-factor for the free exciton in WSe$_2$ is -4.[6,12] However, the plots in Figure 5e and 5f deviate markedly from linearity at $|B| > 2\ T$ (orange-shaded regions), indicating the presence of an additional contribution beyond the conventional Zeeman effect. This kind of asymmetric coupling and splitting that diverges from linearity at high magnetic fields is a well-established signature of the magnetic proximity effect previously observed in AFM/TMD and FM/TMD HSs.[13–15,35,42] The shape of $\Delta E$ as a function of magnetic field can be directly related to the magnetization behavior of a FM material,[13] and thus follows the equation (For more details, see Section 4 in the SI):

(2) $\Delta E(B) = g_{eff} \mu_B \left[ B + B_{exc} \tanh\left(\frac{B}{B_0}\right) \right]$

Here, $B_{exc}$ represent the effective interfacial exchange field, and $B_0$ denotes the characteristic field for saturation transition (i.e., the field required to align most of the magnetic moments in the FM material). As shown in Figures 5e and 5f, this model (black dashed line) provides a better fit to the experimental data than the linear Zeeman expression (red dashed line), particularly at higher magnetic fields. Fitting yields an exchange field of $B_{exc}$=25 T and $B_0$=3 T for HS1, and $B_{exc}$ =27 T with $B_0$=2.9 T for HS2. Exchange fields of similar magnitudes have previously been reported in EuS/WSe$_2$,[13] CrI$_3$/WSe$_2$,[15] and bilayer WS$_2$ positioned on bulk FePS$_3$ slabs.[42] Additional analysis of the Zeeman splitting for the p4 and p5 emission peaks in HS1 is provided in Figure S9 in the SI, showing similar nonlinear behavior and exchange parameters to those extracted for p1, further supporting a common origin for these excitonic peaks.

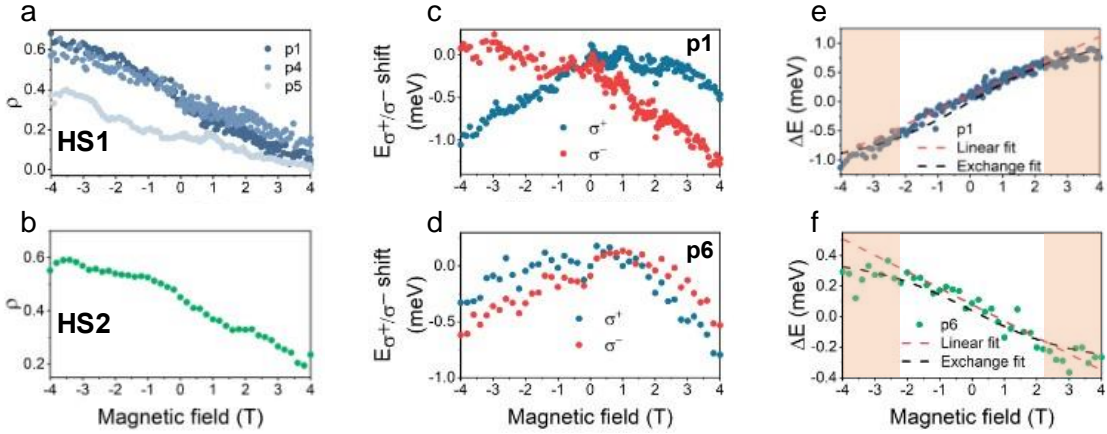

**Figure 5. a, b.** DCP as a function of magnetic field for selected peaks in HS1 (**a**) and HS2 (**b**). **c, d.** Energy shifts of the σ⁺ and σ⁻ components for the p1 peak in HS1 (**c**) and the p6 peak in HS2 (**d**), plotted relative to the zero-field energy of their σ⁺ component. **e, f.** Zeeman splitting of the p1 (**e**) and p6 (**f**) peaks. The red dashed lines are fits to a linear Zeeman model, while the black dashed lines include an additional exchange field term as described by **Eq. 2**.

**2.4. Discussion: The origin of the emitting exciton.** A pivotal question is whether the observed PL emission bands arise from intralayer excitons in WSe$_2$, interlayer excitons, or localized states (e.g., exciton trapped in an interfacial or defect potentials). Several pieces of evidence suggest that the discussed emission events mainly originate from localized intralayer excitons in WSe$_2$ that experience a magnetic proximity effect.

First, the extremely narrow linewidths and the multiplicity of emission peaks strongly suggest the presence of quantum-confined excitonic states. In contrast, a delocalized interlayer exciton would typically manifest as a single, broad emission band - several meV wide - due to its long lifetime (arising from weak oscillator strength) and indirect band nature. Furthermore, in many type-II TMD HSs, interlayer excitons are significantly redshifted by tens to hundreds of meV relative to the bright intralayer exciton.[56] Partial charge transfer in such systems often leads to quenching of the intralayer PL. By contrast, the NiPS$_3$/WSe$_2$ HSs studied here, which also have type-II band alignment, display multiple discrete peaks near the expected energy of the direct intralayer exciton of few-layer WSe$_2$ (around 1.7 eV for trilayer WSe$_2$ at low temperatures).[48,57] The discrete nature of the emission points toward confinement within localized potentials. Indeed, localized defect states in monolayer WSe$_2$ have been observed in the 1.63-1.68 eV range, with emission vanishing above 65 K due to thermal dissociation,[58] which is consistent with the thermal quenching observed in HS1 and HS2, where the PL disappears above 50 K (Figures S4a,b in the SI). However, defect localization is unlikely to be the dominant mechanism in the present case for two reasons: (i) defect states are usually randomly distributed and rarely exhibit reproducible polarization characteristics, and (ii) the observed emission peaks are consistent across different devices and locations. In contrast, moiré-trapped excitons have been widely reported as multiple sharp PL peaks in twisted MoSe$_2$/WSe$_2$ HSs[39] and twisted WSe$_2$ homobilayers.[54] In the latter, for 1.3° twist angle, intralayer excitons in the top layer are trapped by the moiré potential of the bottom layer, yielding multiple peaks between 1.55–1.73 eV,[54] with an energy spread of ~180 meV. A remarkably deep moiré potential, reaching up to 130 meV, has also been reported in MoSe$_2$/MoS$_2$ HSs.[59] Therefore, exciton confinement by a moiré potential is a compelling mechanism in 2D systems. However, in the present work, both NiPS$_3$ and WSe$_2$ are relatively thick, and the two materials are strongly incommensurate, precluding the formation of a well-defined moiré superlattice. Nevertheless, even without a periodic moiré lattice, the interface can host multiple local high-symmetry stacking configurations. Because no long-range commensurate reconstruction occurs, these local registries can arise for a wide range of relative orientations and are therefore not uniquely determined by the twist angle. Such local potentials give rise to spatially varying interface potentials that can act as exciton traps, resembling the confinement observed for moiré excitons in commensurate TMD HSs. This scenario is further supported by the theoretical calculations presented later in the text.

While some charge transfer may still occur, the PL intensity and energy are clearly dominated by WSe$_2$-localized emission. Note that the 1.56 eV peak in HS1 and HS2 falls within the typical range reported for in-plane strained WSe$_2$ emitters.[55,60] However, Raman analysis of the main WSe$_2$ peaks in the heterostructures (Figure S2) reveals no discernible shift, broadening, or splitting, indicating that any residual strain is minimal and unlikely to be the origin of the observed optical properties.

A second key observation is the emergence of spontaneous circular polarization at zero magnetic field (Figure 4a). Pure multi-layered WSe$_2$ emission does not intrinsically exhibit valley polarization. Yet, the observation of 50%-70% DCP in the present case is compelling evidence for exchange coupling between localized WSe$_2$ excitons and NiPS$_3$ spins. A possible origin of the effective exchange field is the presence of uncompensated spins at the NiPS$_3$/WSe$_2$ interface. Such spins may arise from orbital hybridization between NiPS$_3$ and WSe$_2$ or from symmetry breaking at the NiPS$_3$ interface, which can modify the magnetic arrangement. Further support arises from the nonlinear signature observed in the magneto-PL spectra (Figures 5c-f). For isolated WSe$_2$ excitons, the Zeeman splitting is expected to vary linearly with magnetic field, with a characteristic $g$-factor of –4.[6,12] Here, the measured energy splitting ($\Delta E$) deviates from linearity above ~2 T and instead follows Equation 2, which incorporates an additional interfacial exchange field term ($B_{exc}$). This nonlinear Zeeman response closely resembles the behavior recently observed in FePS$_3$/WS$_2$ HSs,[42] reporting a saturation of the valley splitting in few-layers WS$_2$ placed on the AFM material FePS$_3$. In that system, the splitting was likewise attributed to a strong interfacial exchange field, driven by the emergence of FM ordering at the FePS$_3$ interface.

In vdW HSs, interfacial orbital hybridization is expected to play a critical role in determining the electronic and magnetic properties of the combined system.[42,61] Such hybridization is typically dominated by the overlap of out-of-plane orbitals, particularly when energy levels and orbital symmetry are compatible. In the present system, the Ni *3d* orbitals (which have an out-of-plane orientation) can hybridize with the W orbitals across the vdW gap, provided that the local alignment and energy conditions are favorable. The d-d orbital coupling may facilitate strong exchange interactions that not only influence exciton properties but also induce anisotropies and modifications to the local spin structure.[42] Additional evidence for interfacial hybridization and spin rearrangement is provided by the theoretical calculations presented in the following section.

Overall, the evidence suggests that the emitting exciton in NiPS$_3$/WSe$_2$ HSs is an intralayer WSe$_2$ exciton localized by an interfacial potential and subject to an effective exchange field arising from uncompensated spins at the NiPS$_3$ interface. The resulting nonlinear and asymmetric valley splitting is consistent with models of magnetic proximity coupling in vdW HSs.[14,35,39]

2.5. **Discussion: theoretical insights.** Guided by the optical measurements, which indicate that the observed emission originates from the NiPS$_3$/WSe$_2$ interface, the calculations focus on explicitly capturing the interaction between the two materials. To this end, the electronic structure of mono- and bilayer WSe$_2$ interfaced with monolayer NiPS$_3$ was calculated using DFT as implemented in the all-electron FHI-aims code (For more details, see SI, Section 6).[62–64] This reduced model captures the essential interfacial physics while remaining computationally tractable, since the dominant electronic coupling occurs at the interface and layers further away are expected to have a much weaker influence. The resulting band structure for the bilayer WSe$_2$/monolayer NiPS$_3$ HS is shown in Figure 6. The valence band maxima (VBM) are dominated by the WSe$_2$ valence states (black and orange lines in Figure 6), while the conduction band minima (CBM) are composed of Ni *3d* states (green lines in Figure 6). Their distance to the CB edge of WSe$_2$ depends very much on specific details of the experimental setup. For instance, the multilayer WSe$_2$ flakes used in the experiments exhibit a reduced bandgap compared to the bilayer WSe$_2$. In addition, the pronounced linear polarization of the PL emission observed experimentally (Figure 3b,c) provides compelling evidence for broken in-plane symmetry and strongly suggests the existence of a built-in electric field at the NiPS$_3$/WSe$_2$ interface. This internal field is likely a consequence of structural asymmetry, which is inherent to the vertical stacking of the HSs. Additional support for the presence of this field comes from current-voltage (I-V) measurements performed on HS2, which confirms the formation of a p-n junction across the NiPS$_3$/WSe$_2$ interface (see SI, Figure S10). Such an electric field shifts the Ni *3d* states towards the WSe$_2$ conduction band, facilitating the possibility for hybridization or interfacial charge transfer. Therefore, in all further theoretical results, a small external electric field was included, mimicking these experimental peculiarities. A demonstration of the electric field effect is provided in the SI (Figure S12).

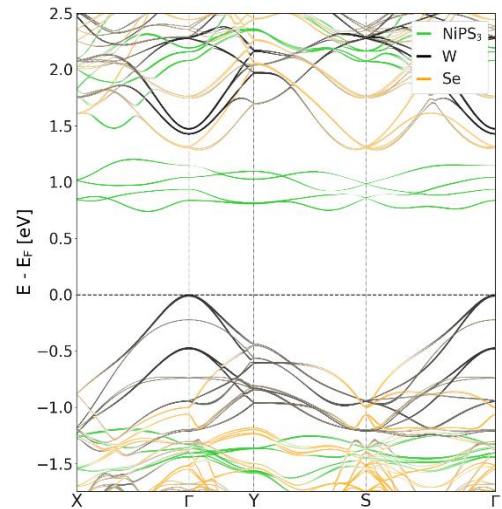

**Figure 6.** Band structure of monolayer NiPS$_3$ on bilayer WSe$_2$. The color indicates the components contributing predominantly to each band. The interfacial dipole leads to a small splitting of the valence-band maxima of the two WSe$_2$ layers. The Brillouin zone and high-symmetry points are shown in Figure 7.

The calculations were initiated with a full relaxation of the HS, including the atomic positions as well as the lattice parameters. The final relaxed structure shows that the lattice parameter of WSe$_2$ increases by approximately 0.4%. Furthermore, the interlayer distance in high-symmetry stackings changes by 0.01 Å, while for smaller twist angles, the change can be twice that value. Although this is smaller than the change of the interlayer distance in TMD HSs, it still indicates that the two systems influence each other on a structural level. To allow for a better comparison with a single monolayer or bilayer WSe$_2$, the lattice of all calculated systems was fixed to the fully relaxed WSe$_2$ bulk lattice.

First, the interface potential was estimated by calculating six high-symmetry stacking configurations, as shown in the SI, Figure S13. The maximal change in the intralayer gap is only about 18.7 meV. Assuming a simple hard sphere model,[65] some of the Se atoms can fit into the corresponding grooves or holes of NiPS$_3$, but not all align perfectly. Accordingly, the interlayer distance is mostly determined by the chalcogen atoms that lie nearly on top of each other; thus, the change in the intralayer gap of WSe$_2$ is much smaller than in TMD HSs.[54,66] The lower interfacial potential estimated by DFT (~18.7 meV) compared to experimental values (40–80 meV) is expected, given DFT's limitations in calculating excited-states energies, exciton binding energies, and the constraints of minimal unit cell simulations that cannot fully capture local structural relaxations.

A direct calculation of the excitonic properties of the relaxed HSs is not feasible due to the large number of electronic states and their complex dispersion as well as the highly correlated nature of the NiPS$_3$ exciton.[67] Yet, the intensity of an optical transition is directly proportional to the squared momentum (or optical) matrix elements (MMEs) between the valence and conduction bands.[68] For TMDs and their HSs, it is known that the strongest interlayer transitions have a two to three orders of magnitude lower intensity than their intralayer counterparts.[68] Comparing the MMEs of the strongest WSe$_2$ intralayer transition with those of the interlayer transitions, a striking difference of at least six orders of magnitude is found. Figure 7a (the left-hand panels) shows the WSe$_2$ intralayer transition with the largest MME, corresponding to the absorption of a circularly polarized photon at the K point (which is backfolded to the $\Gamma$ point). Figure 7b (right-hand panels) displays the strongest interlayer transition at the $\Gamma$ point (i.e., at the lowest energy), corresponding to the absorption of linearly polarized photons. A direct comparison between these two transitions reveals MMEs of $7 \times 10^{-2}$ ($\hbar/a_0$)$^2$ for the WSe$_2$ intralayer absorption versus $2 \times 10^{-8}$ ($\hbar/a_0$)$^2$ for the interlayer absorption. While the PL signal is also influenced by the electron-hole interaction and the increased screening associated with the multilayer character of the samples, these effects impact both the intra- and interlayer excitons. Therefore, it appears highly unlikely that such a large disparity in the MMEs could be compensated, making intralayer transitions the dominant contributor to the observed PL, a conclusion further supported by the experimental observations presented in this work.

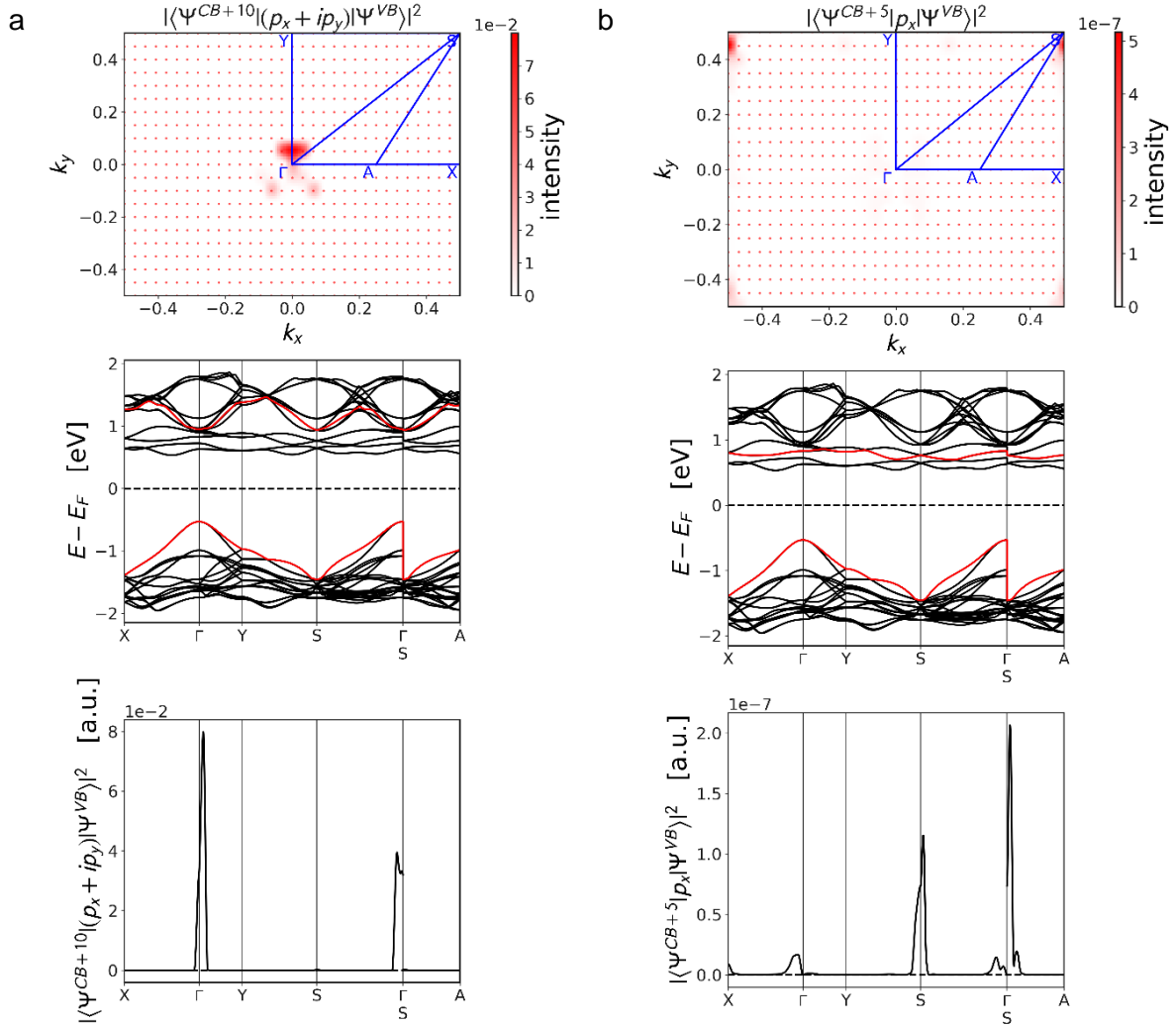

**Figure 7. a.** Intensity of σ$^+$-polarized light for the spin-allowed transition between the VBM and the CBM+1 at the K point of monolayer WSe$_2$ in the monolayer/monolayer HS. The upper, middle, and lower panels show a map of the intensity, the band structure of the HS with the initial and final state marked in red, and the MMEs along the high-symmetry lines indicated in the upper panel, respectively. The K (Q) points of the hexagonal monolayer lattice are mapped (close) to the $\Gamma$ point of the HS, and thus the strongest transition is not to the first CB above the Ni *3d* bands. **b.** Intensity for linearly polarized light for the strongest interlayer transition.

The calculations for the monolayer WSe$_2$/monolayer NiPS$_3$ interface further show an increase in the degree of linear polarization (DLP) to approximately 25% for the strongest intralayer transition at the WSe$_2$ K point, once the NiPS$_3$ bands approach the WSe$_2$ conduction band (see Figure S15 in the SI). Yet, even more importantly, the DLP becomes asymmetric - the difference between the MMEs along X and the Γ-Y increases. This asymmetry becomes even more pronounced in the case of bilayer WSe$_2$ interfaced with monolayer NiPS$_3$ (Figure S16). Additional results and details about the MMEs can be found in the SI, section 6.

Next, the isosurface of the probability density of the final state of the WSe$_2$ intralayer transition, shown in Figure 7a, was calculated and is presented in Figure 8. While this transition has a clear intralayer character, a finite probability density is also observed on the NiPS$_3$ layer, indicating an interlayer hybridization. A similar, though weaker, hybridization is seen in the VBM. This hybridization is further manifested in the preferred spin alignment at the interface, as revealed by the spin texture calculations (presented in the SI, Figure S17). The spin texture analysis shows that the spin character of both the WSe$_2$ conduction bands and the unoccupied Ni 3d bands changes once these states approach each other energetically (e.g., under an interfacial electric field). Pronounced changes appear at and near the Γ point, indicating an interfacial interaction that modifies the preferred spin alignment at the NiPS$_3$/WSe$_2$ interface. Such spin rearrangement provides a microscopic mechanism for the experimentally observed magnetic proximity effect.

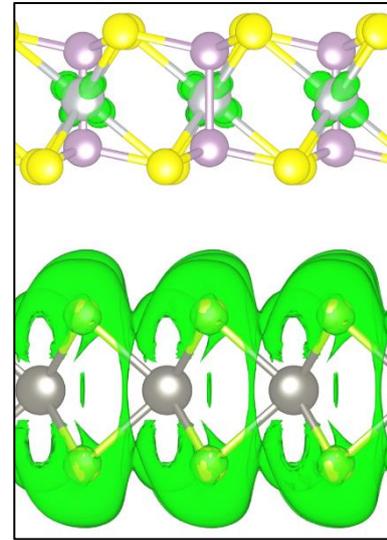

**Figure 8.** Isosurface of the probability density ($|\Psi|^2 = 10^{-4}$) of the wave function for the CBM+10 at the Γ point of monolayer WSe$_2$ (bottom layer) interfaced with monolayer NiPS$_3$ (upper layer), illustrating a finite probability for the state to be on the NiPS$_3$ layer. In the upper layer, yellow, purple, and silver balls represent S, P, and Ni atoms, respectively; in the bottom layer, grey, and yellow balls represent W and Se atoms, respectively.

## 3. Conclusions

This work demonstrates rich excitonic and magnetic phenomena in few-layer NiPS$_3$/WSe$_2$ HSs. Low-temperature μ-PL measurements reveal sharp, discrete emission bands that, together with their strong linear and spontaneous circular polarization, provide compelling evidence that localized intralayer WSe$_2$ excitons confined by an interfacial potential experience a robust magnetic proximity effect arising from uncompensated spins at the NiPS$_3$ interface. The nonlinear Zeeman splitting observed under applied magnetic fields is well described by a model incorporating an interfacial exchange field, with extracted exchange field strengths reaching several tesla. Complementary DFT calculations reveal a varying interfacial potential due to the different atomic registry at the interface, significant interfacial hybridization, modifications to the spin texture, and an enhanced degree of linear polarization when Ni *3d* bands approach the WSe$_2$ conduction bands. These findings underscore the critical role of orbital overlap and spin-valley coupling at the NiPS$_3$/WSe$_2$ interface and establish this system as a promising platform for engineering tunable valleytronic and chiral optoelectronic functionalities in 2D material systems.

**Methods**

*NiPS$_3$.* Bulk NiPS$_3$ single crystals were synthesized via chemical vapor transport synthesis. Briefly, stoichiometric amounts of Ni, red phosphorus, and S$_8$ were ground using a mortar and pestle to form a homogeneous mixture. The resulting powder was transferred to a quartz tube and sealed under vacuum. Then, the tube was placed in a two-zone gradient furnace and heated for one week, with the hot zone (substrate) maintained at 750°C and the cold zone (product) at 690°C. After the growth period, bulk NiPS$_3$ single crystals were collected from the cold zone.

*Device fabrication.* Two types of NiPS$_3$/WSe$_2$ heterostructures (*Device 1* and *Device 2*) were fabricated using a PDMS-based dry transfer method inside a glovebox to prevent contamination and degradation. NiPS$_3$ crystals were synthesized via chemical vapor transport, as described above, while WSe$_2$ and hBN flakes were sourced from HQ Graphene. Si/SiO$_2$ substrates were sequentially cleaned with deionized water, acetone, and isopropanol, followed by drying with nitrogen gas. For *Device 1*, thin flakes of hBN, NiPS$_3$, and WSe$_2$ were mechanically exfoliated and sequentially stacked directly onto the clean Si/SiO$_2$ substrate, with hBN placed first, followed by NiPS$_3$ and then WSe$_2$. For *Device 2*, a Ti/Au (3/27 nm) back electrode was patterned on the Si/SiO$_2$ substrate using electron beam lithography and thermal evaporation prior to stacking. NiPS$_3$ and WSe$_2$ flakes were then transferred in sequence - NiPS$_3$ first, followed by WSe$_2$ - onto the pre-patterned electrode. An hBN flake was transferred on top to encapsulate the stack using a polycarbonate (PC) film at 150 °C. The PC was subsequently removed by rinsing the device in chloroform.

*Raman measurements.* Raman measurements (λ = 532 nm) were performed using a Witec Alpha 300 micro-Raman instrument. The optical pathway was initially calibrated with a standard Si/SiO$_2$ sample, using a 100X (Numerical Aperture 0.9) objective and then the heterostructure sample was measured. Laser power of 5 mW was used with a grating of 1200/500, an integration time of 5 s and 20 accumulations to get one spectrum. Spectra were obtained separately from the NiPS$_3$, WSe$_2$ and the NiPS$_3$/WSe$_2$ regions of the sample.

*P-SHG.* Polarization-resolved second harmonic generation measurements were carried out using a commercial confocal microscope (Witec - Alpha 3000) with a 1064 nm laser and a 100× objective lens (NA 0.9). The laser power was set to 8.54 mW, with a 600 lines/mm grating, 2-second integration time, and five accumulations per spectrum. The incident polarization was rotated from 0° to 360°, and the analyzer was rotated simultaneously with a fixed offset of 90°. All measurements were performed under ambient conditions.

*PL and magneto-PL.* PL and magneto-PL measurements were conducted using a fiber-based confocal microscope integrated in a cryogenic system (attoDRY1000 closed cycle cryostat) equipped with a superconducting magnet and a heater capable of tuning the sample temperature between 5 K – 300 K. Temperature control was maintained by using a Lake Shore PID system. The samples were excited with a 488 nm continuous-wave laser, and the emission was detected using a FERGIE spectrograph. A 495 nm long-pass dichroic mirror, installed on the optical head, was used to direct the laser excitation to the sample while filtering the reflected laser from the emission signal. In addition, a 500 nm long-pass filter was placed in the emission path to further suppress any residual laser signal. For polarization-resolved measurements, a linear polarizer was positioned in the emission path for linear polarization detection, while a linear polarizer combined with a quarter-wave plate (oriented at +45 and -45 degrees to each other's main axis for $\sigma^+$ and $\sigma^-$ detection, respectively) was employed for circular polarization detection.

*DFT.* To calculate the electronic structure of the NiPS$_3$/WSe$_2$ HS systems we used the all-electron FHI-aims code.[63,69,70] The relaxations were performed using the PBE functional[71] on tight tier 1 numeric atom-centered orbitals, including the nonlocal many-body dispersion correction (MBD-nl)[72,73] and scalar relativistic corrections (ZORA), employing a Γ-centered Monkhorst-Pack grids[74] of 30×18×11 for the rectangular supercell shown in Figure S10, which was created using the rectangular magnetic unit cell of NiPS$_3$ in its zigzag magnetic configuration and a rectangular 3×2 supercell of WSe$_2$. This supercell leads to minimal strain in the NiPS$_3$ system ($\approx -1.5\%$, using the WSe$_2$ lattice parameter) while maintaining a reasonably small number of atoms. In the calculations for different stackings, used to estimate the moiré potential, the k-point grids were reduced to 24×16×1. All systems were relaxed until the forces fell below $10^{-4}$ eV Å$^{-1}$, with the in-plane lattice fixed to match the relaxed WSe$_2$ bulk lattice ($a \approx 3.288$ Å). All band structures were calculated using the SCAN[75] functional, including spin-orbit coupling via a post-self-consistent approach[76] and an increased k-point grid of 48×30×1.

**Acknowledgments**

E.L., T.B., T.H. and D.N. thank the support from the Deutsch-Israel Program (DIP, Project No. NA1223/2-1). E.L. thanks the Israel Science Foundation (ISF, Project No. 304/23). T.B. gratefully acknowledge the computing time made available to them on the high-performance computer at the NHR Center of TU Dresden and on the high-performance computers Noctua 2 at the NHR Center PC2. These are funded by the German Federal Ministry of Education and Research and the state governments participating based on the resolutions of the GWK for the national high-performance computing at universities (www.nhr-verein.de/unsere-partner). T.B. thanks James Green from the Molecular Simulations from First Principles e.V. Berlin for fruitful discussions and his help with the spin texture calculations. T.W. acknowledges financial support from National Science Center, Poland under grant no. 2023/48/C/ST3/00309. T.W. gratefully acknowledges Poland's high-performance Infrastructure PLGrid (ACC Cyfronet AGH) for providing computer facilities and support under computational grant no. PLG/2025/018073. N.C. acknowledges the Alexander von Humboldt foundation for the postdoctoral research fellowship. The authors thank Dr. Kusha Sharma for her editing assistance.

**Supporting Information**

Device characterization, Additional PL measurements, reproducibility of the NiPS$_3$/WSe$_2$ HS emission, magnetic proximity fit, I-V measurements, DFT calculations.

# Supporting Information

## Localized Exciton Emission with Spontaneous Circular Polarization in NiPS$_3$/WSe$_2$ Heterostructures

Adi Harchol[1], Shahar Zuri[1], Rajesh Kumar Yadav[2], Nirman Chakraborty[1,3], Idan Cohen[1], Tomasz Woźniak[4], Thomas Brumme[5], Thomas Heine[5,6,7*], Doron Naveh[2*] and Efrat Lifshitz[1*]

8. Schulich Faculty of Chemistry, Solid State Institute, Russell Berrie Nanotechnology Institute, Helen Diller Quantum Information Center, Technion, Haifa 3200003, Israel

9. Faculty of Engineering, Institute for Nanotechnology and Advanced Materials, Bar-Ilan University, Ramat-Gan 52900, Israel.

10. Eduard Zintl Institute of Inorganic and Physical Chemistry, Technical University of Darmstadt, 64287 Darmstadt, Germany

11. University of Warsaw, Faculty of Physics, 02-093 Warsaw, Poland

12. Faculty of Chemistry and Food Chemistry, Technische Universität Dresden, 01062 Dresden, Germany.

13. Helmholtz-Zentrum Dresden-Rossendorf, Center for Advanced Systems Understanding, CASUS, Untermarkt 20, 02826 Görlitz, Germany.

14. Department of Chemistry and ibs for Nanomedicine, Yonsei University, Seodaemun-gu, Seoul, 120-749 South Korea.

1. **Device Characterization**

Figure S1a and S1b display optical images of *Device 1* and *Device 2* (left panels), respectively, alongside atomic force microscopy (AFM) height profiles (right panels) taken along the black dashed lines indicated in the optical images. In *Device 1*, the WSe$_2$ thickness is measured to be ~24.7 nm, corresponding to 35 layers, While in *Device 2*, the WSe$_2$ flake is ~7.2 nm thick, corresponding to 10 layers.[1] The NiPS$_3$ thickness is ~55.4 nm in *Device 1* and ~156.5 nm in *Device 2*, corresponding to ~84 and ~237 layers, respectively.[2]

Figure S1c presents the Raman spectra of *Device 1*, acquired from three different regions: exposed NiPS$_3$ (red curve), exposed WSe$_2$ (blue curve), and the overlapping NiPS$_3$/WSe$_2$ heterostructure region (green curve). The spectra from the exposed NiPS$_3$ and WSe$_2$ regions show good agreement with previously reported Raman signatures of these materials.[2,3] Similar to the observations for *Device 2* described in the main text (Figure 2b), the Raman spectrum recorded from the overlapping NiPS$_3$/WSe$_2$ region is dominated by the characteristic modes of WSe$_2$. In addition, two weaker modes, highlighted with asterisks, are observed and can be attributed to vibrational modes of NiPS$_3$. The presence of both WSe$_2$ and NiPS$_3$ Raman features confirms the successful assembly of the heterostructure in this region.

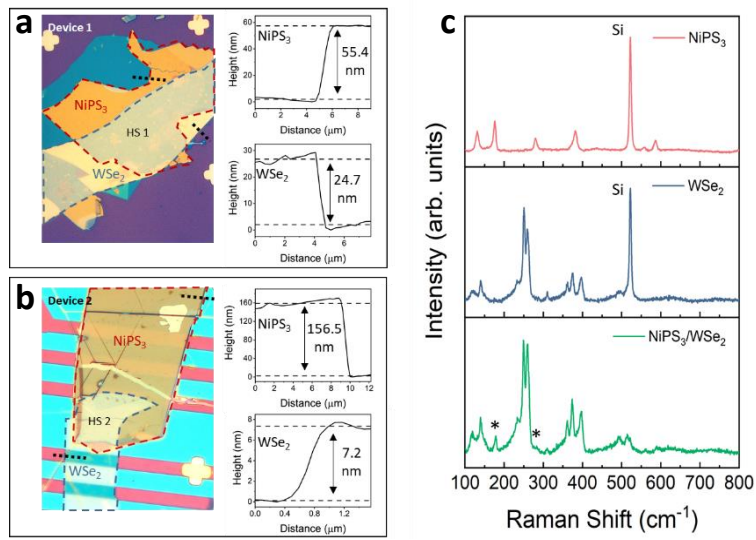

**Figure S1. a, b.** Optical images of *Device 1* (**a**, left) and *Device 2* (**b**, left), each with a corresponding AFM height profile (right) taken along the black dashed line in the optical image. **c.** Raman spectra of *Device 1* recorded at three different locations on top of the sample: exposed NiPS$_3$, exposed WSe$_2$, and NiPS$_3$/WSe$_2$ heterostructure (from top to bottom), measured under 532 nm laser excitation at room temperature.

1.1. **Strain Analysis by Raman Spectroscopy**

The absence of strain in the heterostructures was further confirmed through Raman spectroscopy by showing that the primary vibrational modes of WSe$_2$ in the heterostructure regions do not deviate from those measured in the exposed WSe$_2$ areas and are consistent with previously reported values for unstrained crystals. Raman spectra acquired from both the exposed WSe$_2$ and the NiPS$_3$/WSe$_2$ regions in *Devices 1* and *2* were fitted using Lorentzian functions (Figure S2), and the extracted peak positions are summarized in Table S1. The results reveal that the Raman peaks of WSe$_2$ in the heterostructures exhibit no appreciable spectral shifts relative to their exposed counterparts. In addition, the narrow linewidths observed in all spectra are indicative of high crystalline quality and minimal disorder.

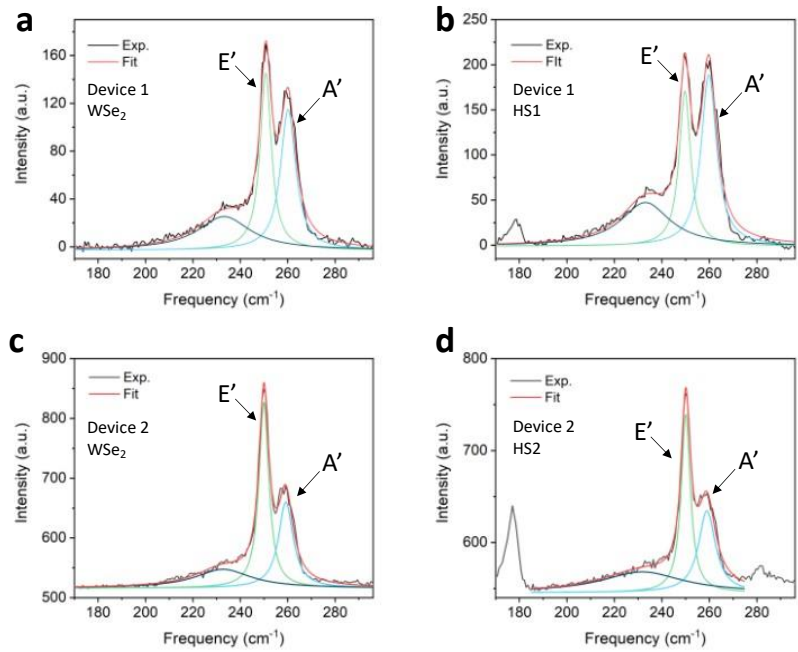

**Figure S2.** Fitted Raman spectra of the **a,c.** exposed WSe$_2$ region and the **b,d.** NiPS$_3$/WSe$_2$ region for *Device 1* **(a,b)** and *Device 2* **(c,d)**.

**Table S1.** Extracted peak positions from Lorentzian-fitted Raman spectra of the exposed WSe$_2$ and the NiPS$_3$/WSe$_2$ regions, in *Device 1* and *Device 2*. Reported values for unstrained WSe$_2$ is also included for comparison.

| Raman mode | Reported frequency (cm$^{-1}$)* | This work *Device 1* | | This work *Device 2* | |
|---|---|---|---|---|---|
| | | WSe$_2$ | HS1 | WSe$_2$ | HS2 |
| $E'$ | 250 | 250.6 | 249.6 | 249.9 | 250 |
| $A'_1$ | 260 | 260.1 | 259.7 | 259.2 | 258.5 |

*Ref.[4]

### 1.2. Polarized Second Harmonic Generation (P-SHG) Fitting

The fitting expression for SHG in WSe$_2$ is determined by its non-centrosymmetric crystal structure and threefold rotational symmetry. For unstrained WSe$_2$, the SHG intensity as a function of polarization angle ($\theta$) follows the equation:[5]

$$I(2\omega, \theta) = \cos^2(3\theta - \phi)$$

Where $\phi$ is the phase offset, reflecting crystal orientation.

In contrast, NiPS$_3$ belongs to the centrosymmetric point group *2/m*, and its SHG response arises from electric quadrupole contributions. The fitting expression, adopted from Ref.[6] for parallel polarization geometry, is:

$$I(2\omega, \theta) \propto \left[\left(\chi_{xyzy} + \chi_{yxzy} + \chi_{yyzx}\right) \cos^2\theta \sin\theta + \chi_{xxzx} \sin^3\theta\right]^2$$

Where $\chi_{ijkl}$ are components of the electric quadrupole susceptibility tensor associated with the *2/m* crystallographic symmetry of NiPS$_3$.

## 2. Additional PL Measurements
### 2.1. PL Measurement of WSe$_2$

The PL of the exposed WSe$_2$ in *Device 1* was measured under identical conditions as the HS1 measurements (4 K and same laser power). As shown in Figure S3, no detectable PL signal was observed, confirming that the emission from the HSs originates from the interface.

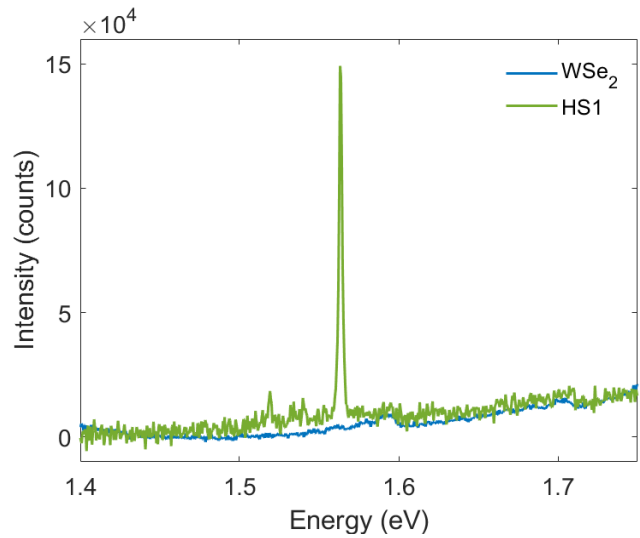

**Figure S3.** PL measurements of the exposed WSe$_2$ (blue) and the HS1 (green) regions at 4 K.

## 2.2. Temperature-Dependent Photoluminescent Measurements of HS1 and HS2

Figures S4a and S4b show color maps of the temperature-dependent photoluminescent (PL) spectra of HS1 and HS2, respectively. In both heterostructures, the PL is strongly quenched at temperatures above 50 K. The dominant emission line at 1.56 eV (p1 in HS1 and p6 in HS2) was fitted at each temperature using a Gaussian line-shape and the extracted parameters - peak energy, full width at half maximum (FWHM), and integrated intensity - are plotted in Figures S4c and S4d for HS1 and HS2, respectively. The resulting trends were well described by established models for exciton behavior in semiconductors (red curves in Figures S4c,d).

The PL intensity as a function of temperature follows the Arrhenius equation:[7]

$$I(T) = \frac{I_0}{1 + a \cdot \exp\left(-\frac{E_a}{k_B T}\right)}$$

where $I_0$ is the low-temperature PL intensity, $a$ is the ratio of radiative to non-radiative decay rates, $E_a$ is the activation energy, and $k_B$ is the Boltzmann constant.

The temperature-dependent of the FWHM is governed by exciton-phonon interactions and follows:[8]

$$FWHM(T) = \Gamma_0 + \frac{\Gamma_{LO}}{\exp\left(\frac{E_{LO}}{k_B T}\right) - 1}$$

where $\Gamma_0$ is the broadening at 0 K, and $\Gamma_{LO}$ reflects the broadening produced by optical phonons with energy $E_{LO}$.

Finally, the temperature-dependent shift of the emission energy is described by the Varshni relation:[9]

$$E_{gap}(T) = E_0 - \frac{\alpha T^2}{T + \beta}$$

where $E_0$ is zero-temperature bandgap, and $\alpha$ and $\beta$ are constants. The extracted fitting parameters for both heterostructures are summarized in Table S2.

**Table S2.** Extracted fitting parameters from temperature-dependent PL analysis of HS1 and HS2, based on exciton models for exciton models for intensity quenching, linewidth broadening, and bandgap shift.

|  | Fit parameter | p1 (HS1) | p6 (HS2) |
|---|---|---|---|
| **Integrated PL intensity** | $a$ | 89.8±24.4 | 449.7±91.8 |
|  | $E_a$ | 15.1±1.0 [meV] | 16.9±0.6 [meV] |
| **FWHM** | $\Gamma_0$ | 2.2±0.5 [meV] | 1.8±0.2 [meV] |
|  | $\Gamma_{LO}$ | 9.5±1.9 [meV] | 4.9±0.6 [meV] |
|  | $E_{LO}$ | 5.9±0.7 [meV] | 4.3±0.3 [meV] |
| **Energy shift** | $\alpha$ | 1.1x10$^{-3}$±0.1 x10$^{-3}$ [eV·T$^{-1}$] | 4.3x10$^{-4}$±0.3 x10$^{-4}$ [eV·T$^{-1}$] |
|  | $\beta$ | 388±50 [T] | 400±31 [T] |

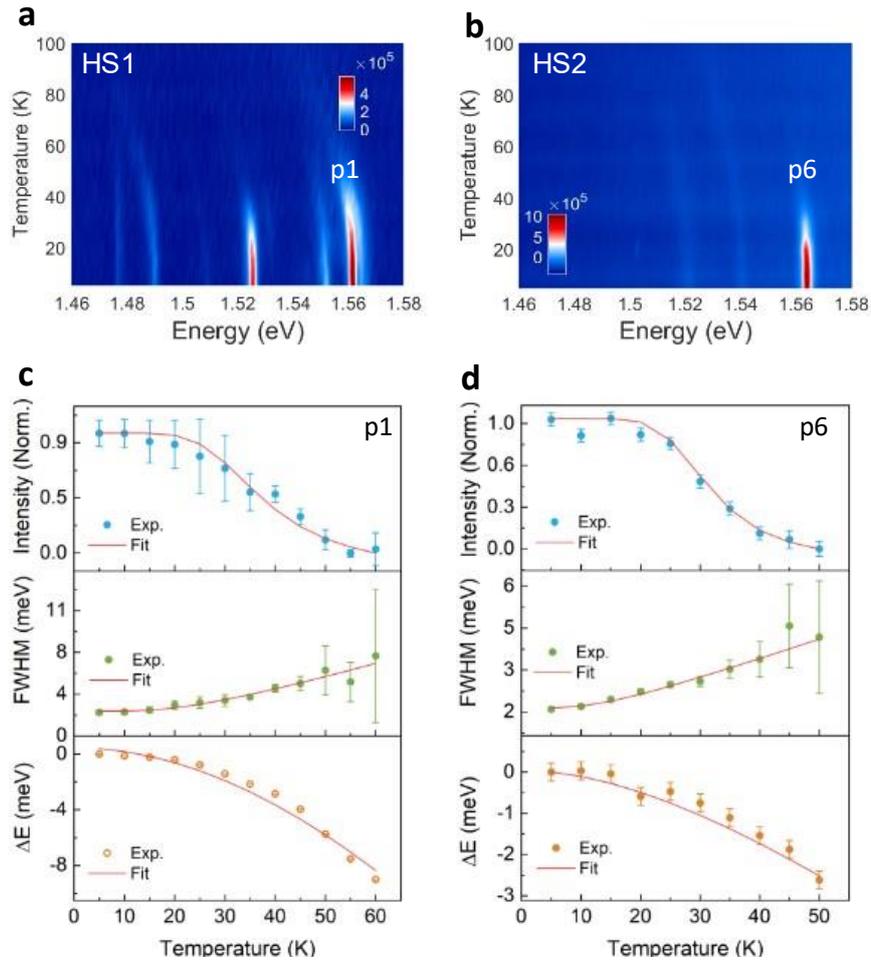

**Figure S4.** Temperature-dependent PL color map of **a.** HS1 and **b.** HS2, measured under 480 nm laser excitation. Panels **c**, and **d** show, from top to bottom, the integrated PL intensity, FWHM, and emission energy shift as a function of temperature for peaks p1 and p6, respectively, extracted from Gaussian fits to the temperature-dependent spectra. The red lines represent fits based on exciton behavior models in semiconductors.

### 2.3. Linear Polarization of HS1

Figure S5a,b present PL spectra of HS1 at different polarizer angles. As seen in Figure S5a, the emission peaks in the HS exhibit pronounced linear polarization. Figure S5b focuses on the spectral region near the NiPS$_3$ exciton at 1.476 eV. No clear polarization dependance is observed for this peak, which we attribute to its weak emission signal in the heterostructure, rendering it indistinguishable from the background noise.

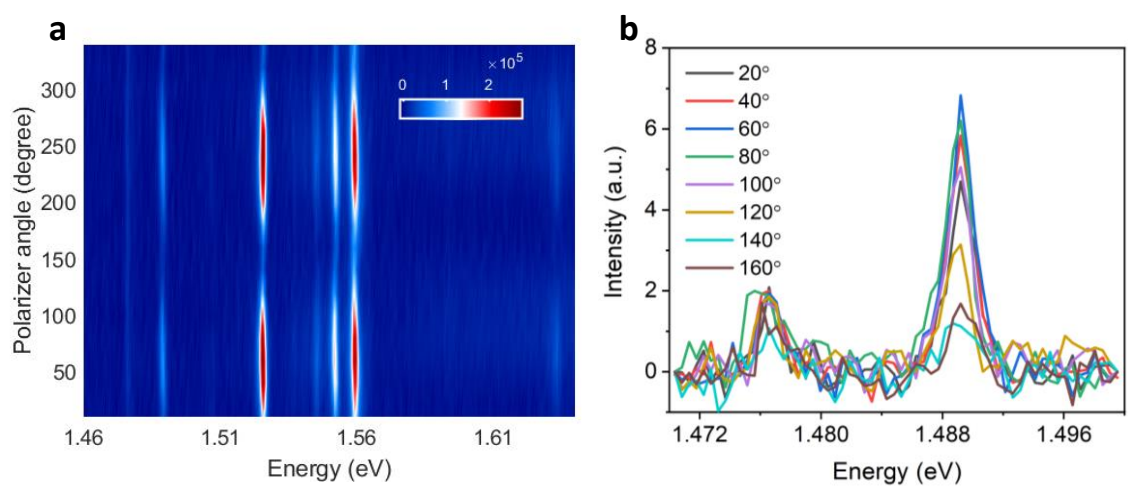

**Figure S5. a.** Polarization-resolved PL color map of HS1. **b.** PL spectra of HS1 at different polarizer angles, focusing on the spectral window around 1.476 eV.

### 2.4. Calibration of the circular polarization detection

To verify that the circular polarization detected in the NiPS$_3$/WSe$_2$ heterostructures does not originate from instrumental artifacts, a control measurement was performed on bare NiPS$_3$. The PL emission of NiPS$_3$ is known to exhibit strong linear polarization while showing negligible circular polarization. Figure S6 presents the circularly resolved PL spectra measured from the bare NiPS$_3$ region under the same optical configuration used for the heterostructures. The σ$^+$ and σ$^-$ components overlap, resulting in a negligible degree of circular polarization. This result confirms that the polarization detection path does not introduce artificial circular polarization and validates the reliability of the circular polarization measurements presented in the main text.

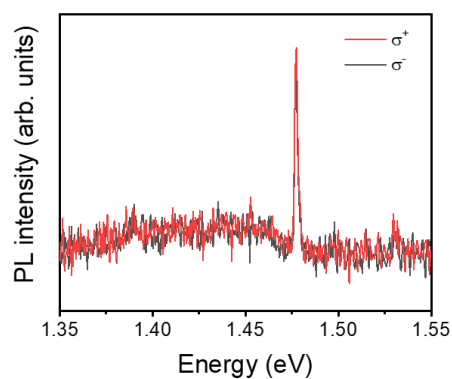

**Figure S6.** Circularly resolved PL spectra measured from bare NiPS$_3$.

## 2.5. Circularly Polarized PL at Different Locations for HS1 and HS2

Figure S7a and S7b show circularly polarized PL spectra acquired at different locations and on different days for HS1 and HS2, respectively. The emission spectra are relatively uniform across all spots, with small variations in peak intensity. In addition, the degree of circular polarization is preserved across different measurement spots, indicating that the emission does not originate from localized defects.

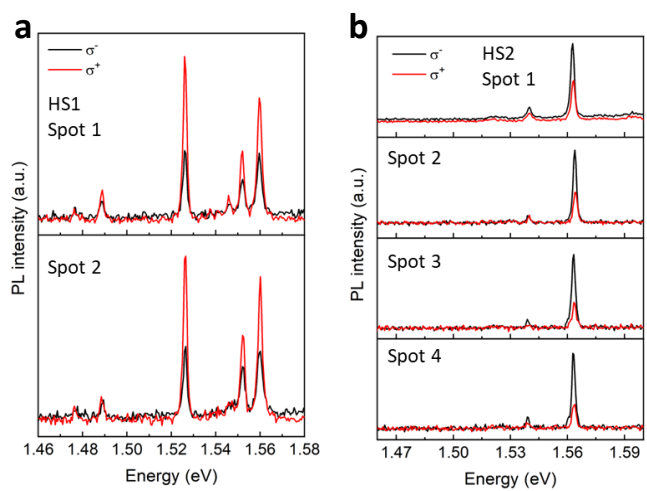

**Figure S7.** Circularly polarized PL spectra of **a.** HS1 and **b.** HS2 acquired at different days and spots on top of the HSs.

### 3. Reproducibility of the NiPS$_3$/WSe$_2$ HS Emission – HS3 and HS4

**Figure S8a and S8b show optical microscope images of two additional NiPS$_3$/WSe$_2$ heterostructure, HS3 and HS4, respectively. The PL spectrum of HS3, acquired at 5 K (Figure S8b), reveals a sharp emission line at 1.56 eV and an additional peak at 1.476 eV attributed to NiPS$_3$. This 1.56 eV emission line is quenched above 30 K, as shown in Figure S8e. Figure S8d presents polarization-resolved PL spectra of HS4 at 5 K and 0 T, showing the same emission at 1.56 eV with substantial DCP (~40%) at zero magnetic field. This emission is also quenched above 30 K, as demonstrated in Figure S8f. The presence of this strong 1.56 eV peak, nonzero DCP at B=0 T, and its temperature-dependent trend across multiple independent devices (HS1-HS4) further supports the reproducibility and reliability of the optical response in the NiPS$_3$/WSe$_2$ heterostructures presented in this work.**

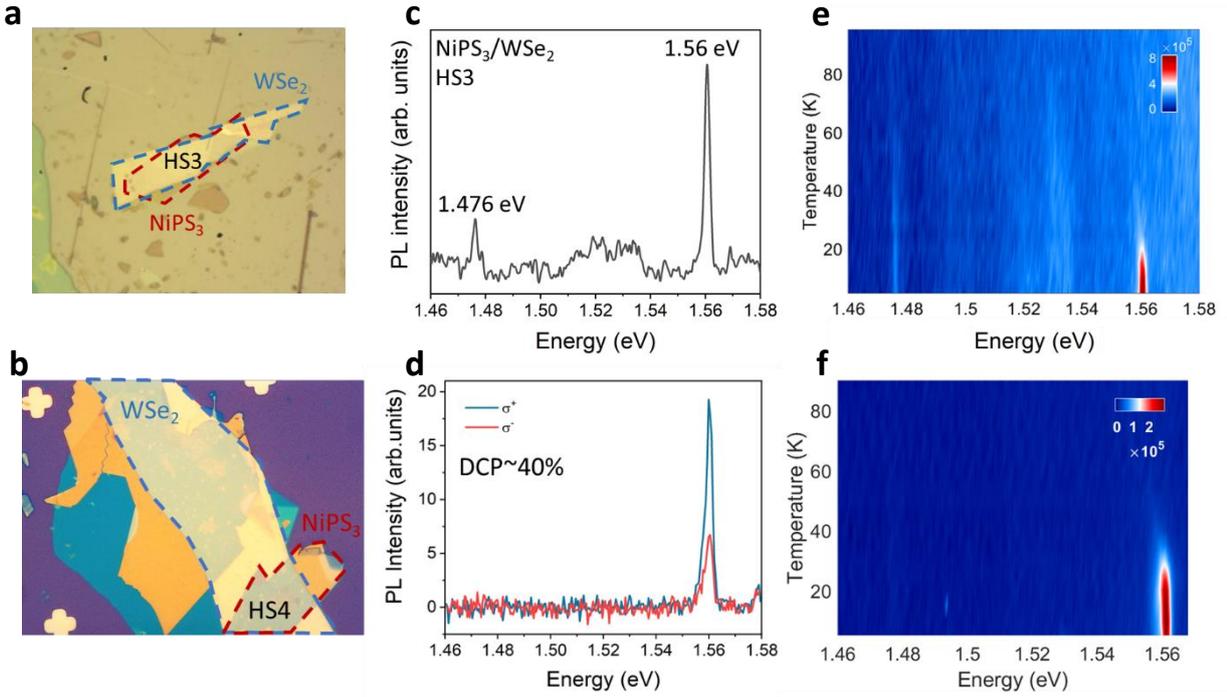

**Figure S8.** Optical microscope images of **a.** HS3 and **b.** HS4. The red and blue dashed lines indicate the boundaries of the NiPS$_3$ and WSe$_2$ flakes, respectively. PL spectra of the overlapping NiPS$_3$/WSe$_2$ region in **c.** HS3 and **d.** HS4 acquired at 5 K under 480 nm laser excitation. For HS4 (**d**), the PL was measured under circular polarization detection at 0 T. Temperature-dependent PL color maps of **e.** HS3 and **f.** HS4.

### 4. Magnetic Proximity Fit

For a ferromagnet with localized spin-½ ensemble in an external magnetic field *B* at temperature *T*, the equilibrium magnetization, $M(B,T)$, is given by:[10]

(1) $M(B,T) = N\mu_B \tanh\left(\frac{\mu_B B}{k_B T}\right) = M_s \tanh\left(\frac{B}{B_0}\right)$

Where *N* is the total number of spin moments in the sample, $\mu_B$ is the Bohr magneton, and $k_B$ is the Boltzmann constant.

Here,

$M_s = N\mu_B$ is the saturation magnetization.

$B_0 = \frac{k_B T}{\mu_B}$ is the characteristic field scale set by the thermal fluctuations.

**Equation 1** reflects the behavior where the magnetization increases linearly with B at low fields and saturates at high fields.

When magnetic proximity effects are considered, an effective exchange field $(B_{ex}^*)$ proportional to the substrate magnetization is introduced,[11,12] such that:

(2) $B_{ex}^* = B_{exc} \tanh\left(\frac{B}{B_0}\right)$

Where $B_{exc}$ is the saturated exchange field, analogous to $M_s$ for the magnetization.

Thus, the total effective field acting on an exciton in the 2D material is the sum of the external field and the exchange field contribution:

(3) $B_{tot}(B) = B + B_{ex}^* = B + B_{exc} \tan\left(\frac{B}{B_0}\right)$

Accordingly, for carriers with an effective *g*-factor, $g_{eff}$, the total Zeeman splitting $\Delta E(B)$, becomes:

(4) $\Delta E(B) = g_{eff} \mu_B B_{tot} \rightarrow \Delta E(B) = g_{eff} \mu_B \left[B + B_{exc} \tan\left(\frac{B}{B_0}\right)\right]$

### 4.1. Exchange Fit for p4 and p5 Peaks in HS1

Figure S9 presents the energy splitting of the p4 (top) and p5 (bottom) exciton peaks in HS1 as a function of magnetic field. Both features exhibit nonlinear Zeeman behavior similar to that observed for the p1 peak. The energy splitting was fitted using the exchange model described by Equation (2) in the main text, and the extracted parameters $B_{exc}$ and $B_0$ are indicated within the plots. The similarity of these parameters to those extracted for p1 supports a common origin for all three excitonic states.

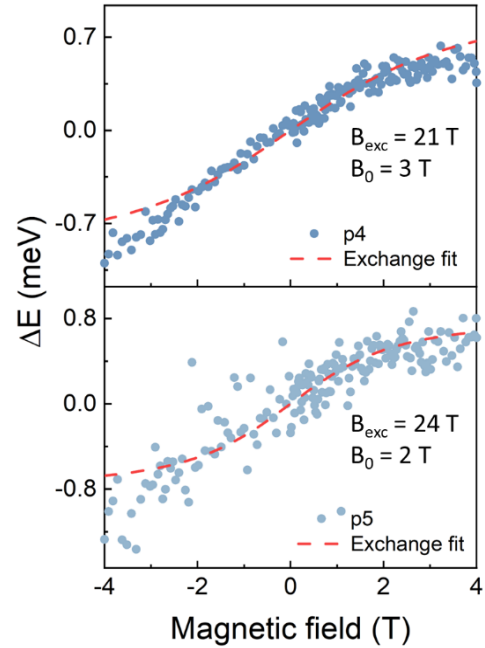

**Figure S9.** Energy splitting of the p4 (top) and p5 (bottom) peaks in HS1 as a function of magnetic field, along with fits to the exchange equation described in the main text.

5. **I-V measurements**

Current-voltage (I-V) measurements were performed at room temperature under vacuum conditions (~$10^{-5}$ Torr) using a Keysight B2919 source-measure unit. Figure S10 presents the I-V measurement of the NiPS$_3$/WSe$_2$ heterostructure in *Device 2*, exhibiting a clear rectifying behavior - characterized by suppressed current under reverse bias and nonlinear increase under forward bias. This asymmetry is consistent with the formation of a p-n junction at the NiPS$_3$/WSe$_2$ interface and supports the existence of a built-in electric field.

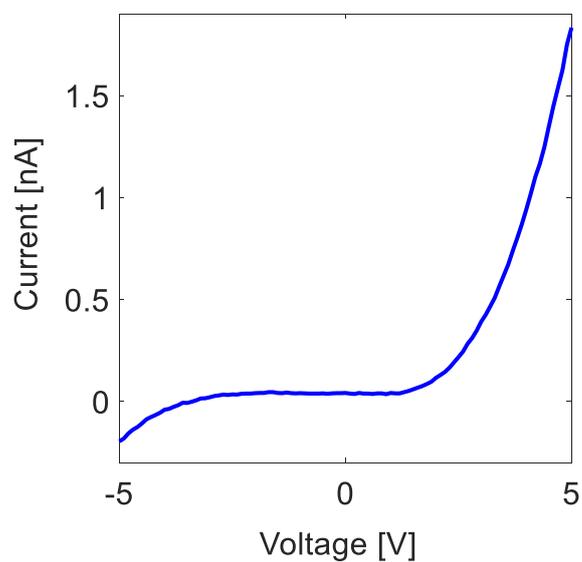

**Figure S10.** Current-voltage characteristics of the NiPS$_3$/WSe$_2$ heterostructure in *Device 2*, measured at room temperature under vacuum.

## 6. Density-Function Theory Calculations

In order to calculate the electronic structure of the NiPS$_3$/WSe$_2$ heterostructure systems we used the all-electron FHI-aims code.[13–15] The relaxations were performed using the PBE functional[16] on tight tier 1 numeric atom-centered orbitals, including the nonlocal many-body dispersion correction (MBD-nl)[17,18] and scalar relativistic corrections (ZORA), employing Γ-centered Monkhorst-Pack grids[19] of 30×18×11 for the rectangular supercell shown in Figure S11, which was created using the rectangular magnetic unit cell of NiPS$_3$ in its zigzag magnetic configuration and a rectangular 3×2 supercell of WSe$_2$. This supercell leads to minimal strain in the NiPS$_3$ system ($\approx -1.5\%$, using the WSe$_2$ lattice parameter) while maintaining a reasonably small number of atoms. In the calculations for different stackings, used to estimate the moiré potential, the k-point grids were reduced to 24×16×1. All systems were relaxed until the forces fell below $10^{-4}$ eV Å$^{-1}$, with the in-plane lattice fixed to match the relaxed WSe$_2$ bulk lattice ($a \approx 3.288$ Å). All band structures were calculated using the SCAN[20] functional, including spin-orbit coupling via a post-self-consistent approach[21] and an increased k-point grid of 48×30×1.

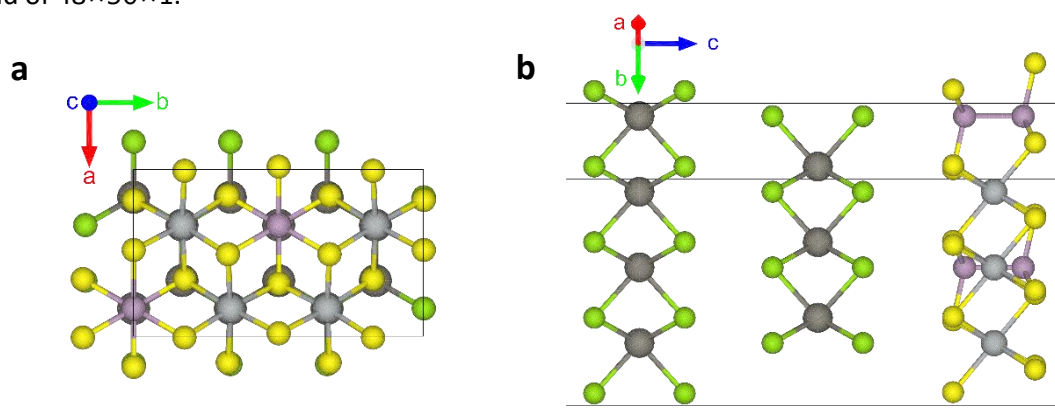

**Figure S11.** Top-view (**a**) and side-view (**b**) of the bilayer WSe$_2$/monolayer NiPS$_3$ heterostructure. Green, dark grey, light grey, rose, and yellow spheres represent Se, W, Ni, P, and S atoms, respectively. The unit cell is indicated by grey lines, and the directions of the lattice vectors are shown in the upper left corner of each panel.

Figure S12 shows the resulting band structures for the monolayer WSe$_2$/monolayer NiPS$_3$ and bilayer WSe$_2$/monolayer NiPS$_3$ systems, both with and without external electric field. One can clearly see that under small negative fields, the Ni *3d* states shifts toward the WSe$_2$ conduction band, leading to hybridization/overlap between the Ni *3d* states and the WSe$_2$ conduction bands. This effect also occurs in the bilayer WSe$_2$/monolayer NiPS$_3$ system; however, an additional splitting appears, coming from the different potential felt by the layers of WSe$_2$. The hybridization complicates the comparison of the 2D plots of the momentum matrix elements (MMEs), which are always calculated between one specific pair of initial and final states. In the monolayer/monolayer system, for example, the tungsten-based conduction band minimum + 1 (CBM+1) at the Γ point (corresponding to the backfolded K point of the unit cell) actually crosses other bands in the vicinity of the Γ point. Thus, the highest contributions for the MMEs are expected for spin-allowed transition from the same initial state (valence band maximum (VBM) – #1216) to a set of different final states. Yet, before discussing the MMEs and the degree of circular and linear polarization (DCP/DLP), we disscuss the interface potential in Figure S13.

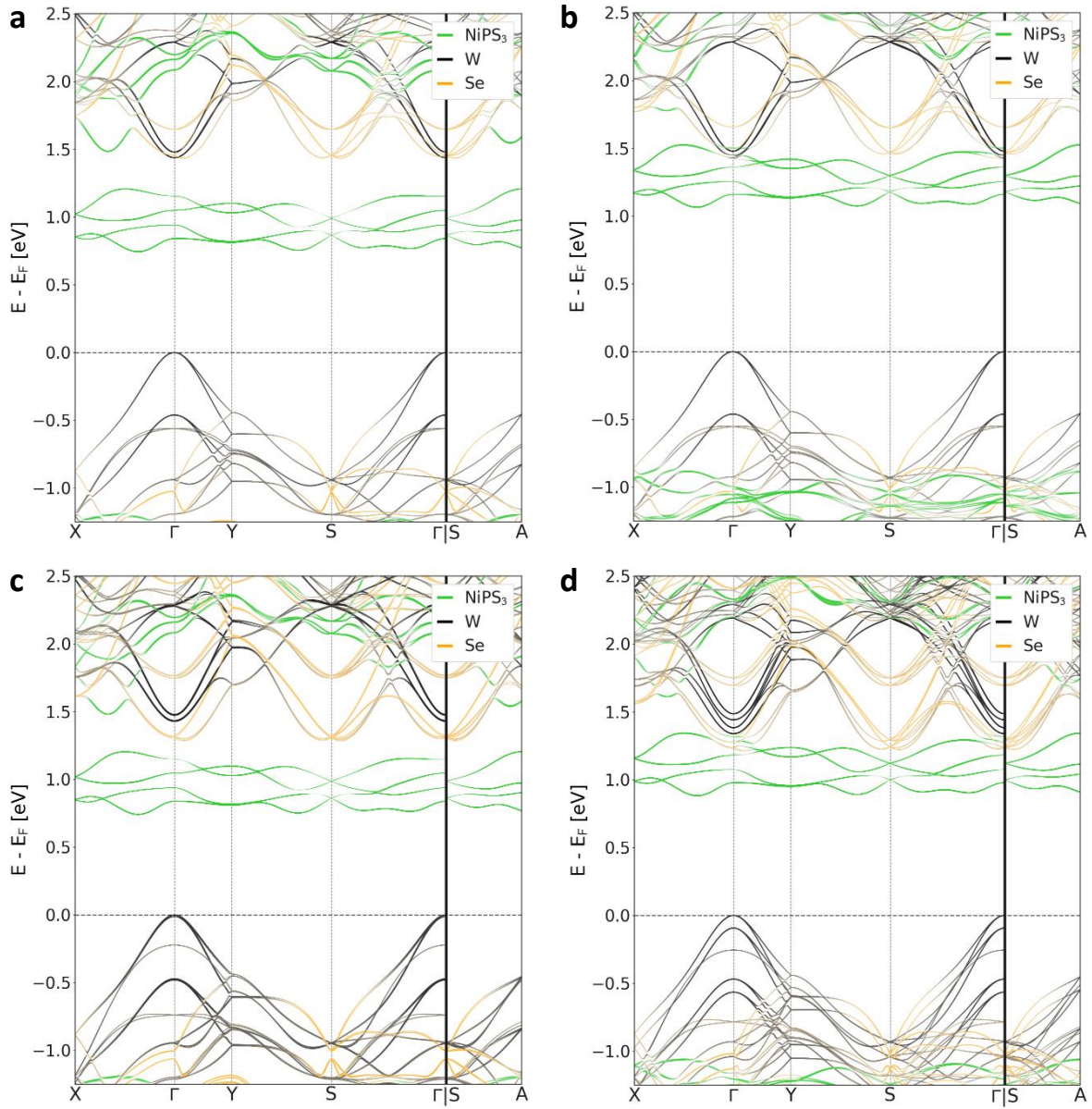

**Figure S12.** Band structures of the monolayer WSe$_2$/monolayer NiPS$_3$ heterostructure **a.** without external electric field and **b.** with a finite electric field of -0.25 eV Å$^{-1}$, and the bilayer WSe$_2$/monolayer NiPS$_3$ heterostructure **c.** without external electric field and **d.** with a finite electric field of -0.10 eV Å$^{-1}$. In panel **b** and **d**, the Ni *3d* states are shifted towards the WSe$_2$ conduction band, leading to hybridization and overlap of the states. In the bilayer/monolayer system (panel **d**), an additional splitting appears, coming from the different potential/field felt by the layers of WSe$_2$.

The interface potential shown in Figure S13 was calculated for the lowest intralayer WSe$_2$ transition without an additional external electric field, corresponding to the system presented in Figure S12c. Since the NiPS$_3$ lattice parameter is about twice that of WSe$_2$, the two layers do not align perfectly, and there is always one Se atom nearly positioned on-top an S atom. As a result, the structural changes are minor, and the small variations observed in the intralayer gap of WSe$_2$ are mainly attributed to changes in hybridization. The application of a small finite electric field does not significantly affect this gap, as can also be seen by comparing the left-hand with the right-hand panels in Figure S12.

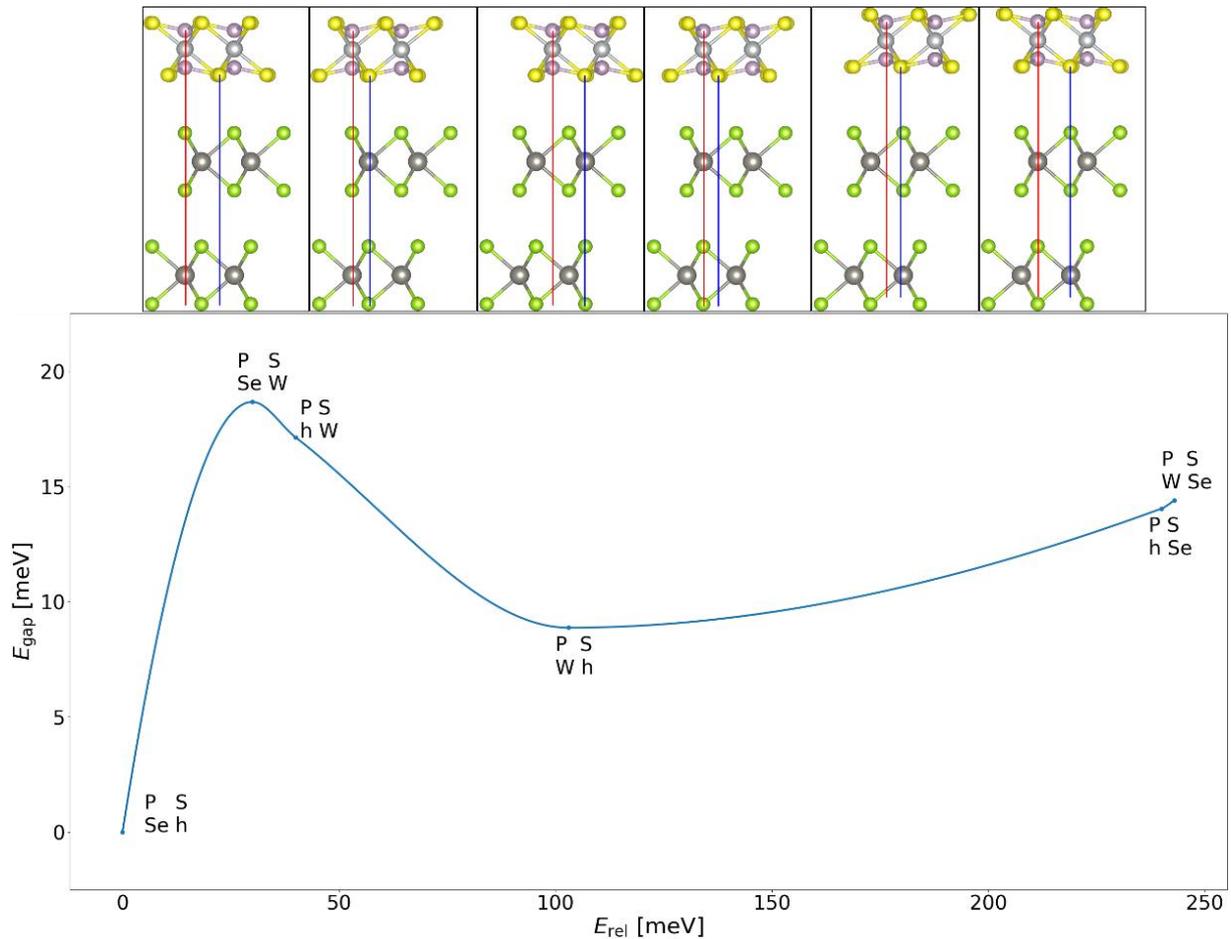

**Figure S13.** Different stacking configurations of monolayer NiPS$_3$ on bilayer WSe$_2$ (upper panel) and the corresponding change in the energy of the lowest WSe$_2$ intralayer transition at the K point (bottom panel), i.e., the change in the direct bandgap, $E_{gap}$. The bandgap is plotted with respect to the relative total energy of the corresponding configuration, $E_{rel}$. The stacking geometries are sketched above the graph, where the length of the red and blue lines are kept constant, allowing for easier comparison of the interlayer distances between different stackings. The stacking nomenclature is inspired by those of the TMD HSs: the upper two letters correspond to two high-symmetry positions in NiPS$_3$, and the lower two letters to high-symmetry positions in the interface WSe$_2$ layer, with "h" indicating the hollow site. The structural images are arranged in the same order as in the graph below.

As already mentioned above, the addition of electric field (mimicking the experimental conditions), leads to the crossing of multiple bands near the $\Gamma$ point, thereby complicating a direct comparison of the MMEs for a single transition with experimental results. As illustrated in Figure S14, multiple WSe$_2$ intralayer transitions originating from the VBM as the initial state exhibit comparable MME strength (CBM+9 vs. CBM+11). Consequently, akin to the experimental scenario where all transitions contribute to the optical response, it is in principle necessary to consider all relevant transitions over a fine k-point grid.

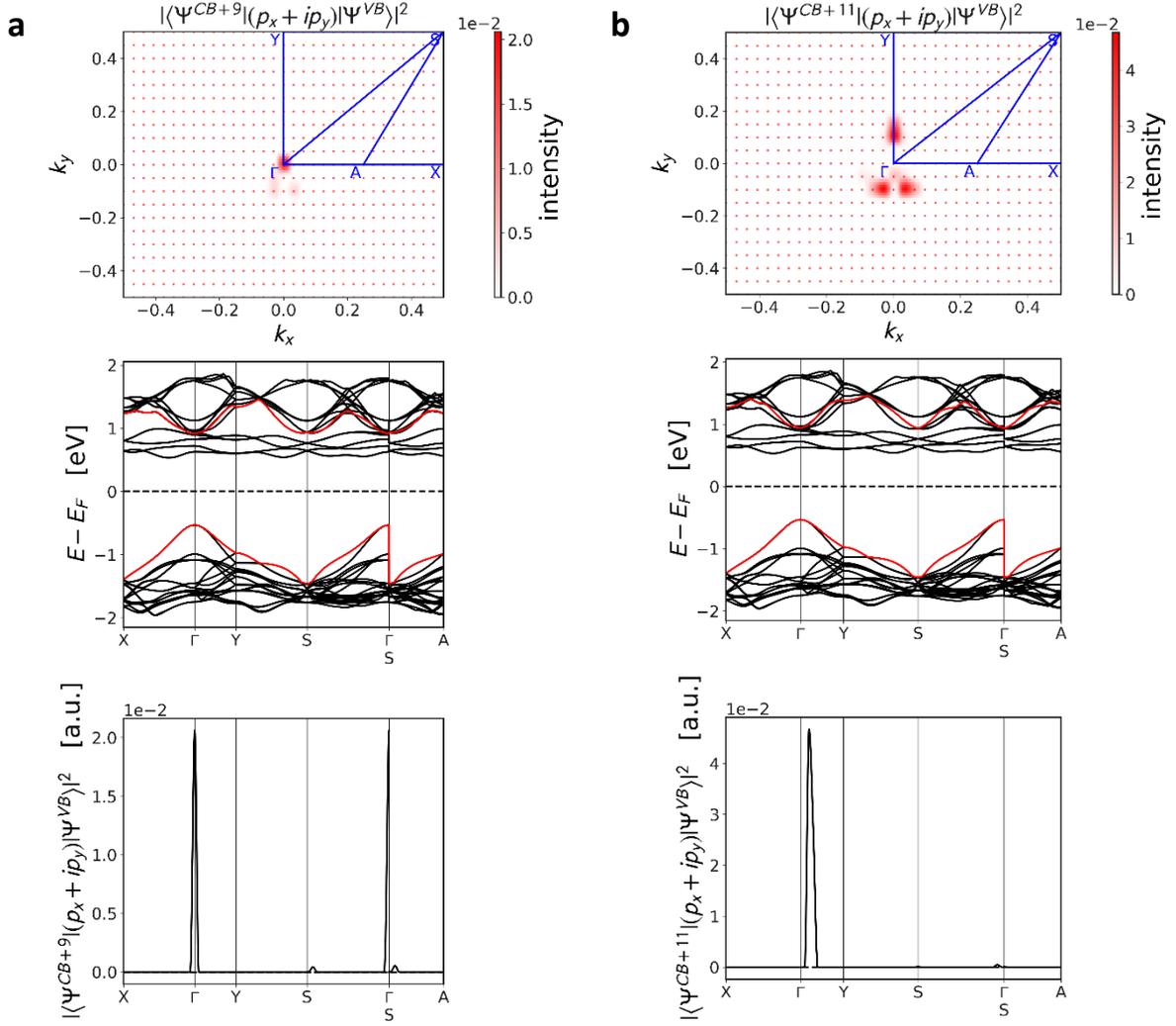

**Figure S14.** MMEs for circularly polarized light of the strongest WSe$_2$ intralayer transition with the VB as initial state: **a.** to the CBM+9 **b.** to CBM+11. The transition to the CB+10 is shown in the main text. The transition VB to CB+12 has similar strength but the hot spots are further away from the Γ point.

Next, we compared the DLP for the strongest WSe$_2$ intralayer transition in Figure S15 without and with external electric field, mimicking the experimental conditions (see discussion in the main text). As the conduction bands of the two systems approach one another, the DLP increases, and the texture in the 2D map also changes, as shown in the top panels of Figures S15. Similar effects can also be observed in the bilayer WSe$_2$/monolayer NiPS$_3$ heterostructure, as shown in Figure S16.

Finally, the spin textures of the WSe$_2$ and NiPS$_3$ conduction bands are compared to examine how they change as the states approach each other. Figure S17 displays the spin textures for the different systems, showing the spin expectation values for the bands. The z-component $\langle\sigma_z\rangle$ is indicated by the color code of the arrows, in which green typically marks spin-degenerate bands. The arrow directions represent the in-plane spin components, with $\langle\sigma_x\rangle$ and $\langle\sigma_y\rangle$ align along the momentum (**k**) and energy directions, respectively. Note that these visualizations are qualitative, as the results are interpolated from a **k**-grid with increased density near the conduction-band minima at Γ/S (containing more than 11,000 k points), and an additional scaling factor is used to enhance visibility.

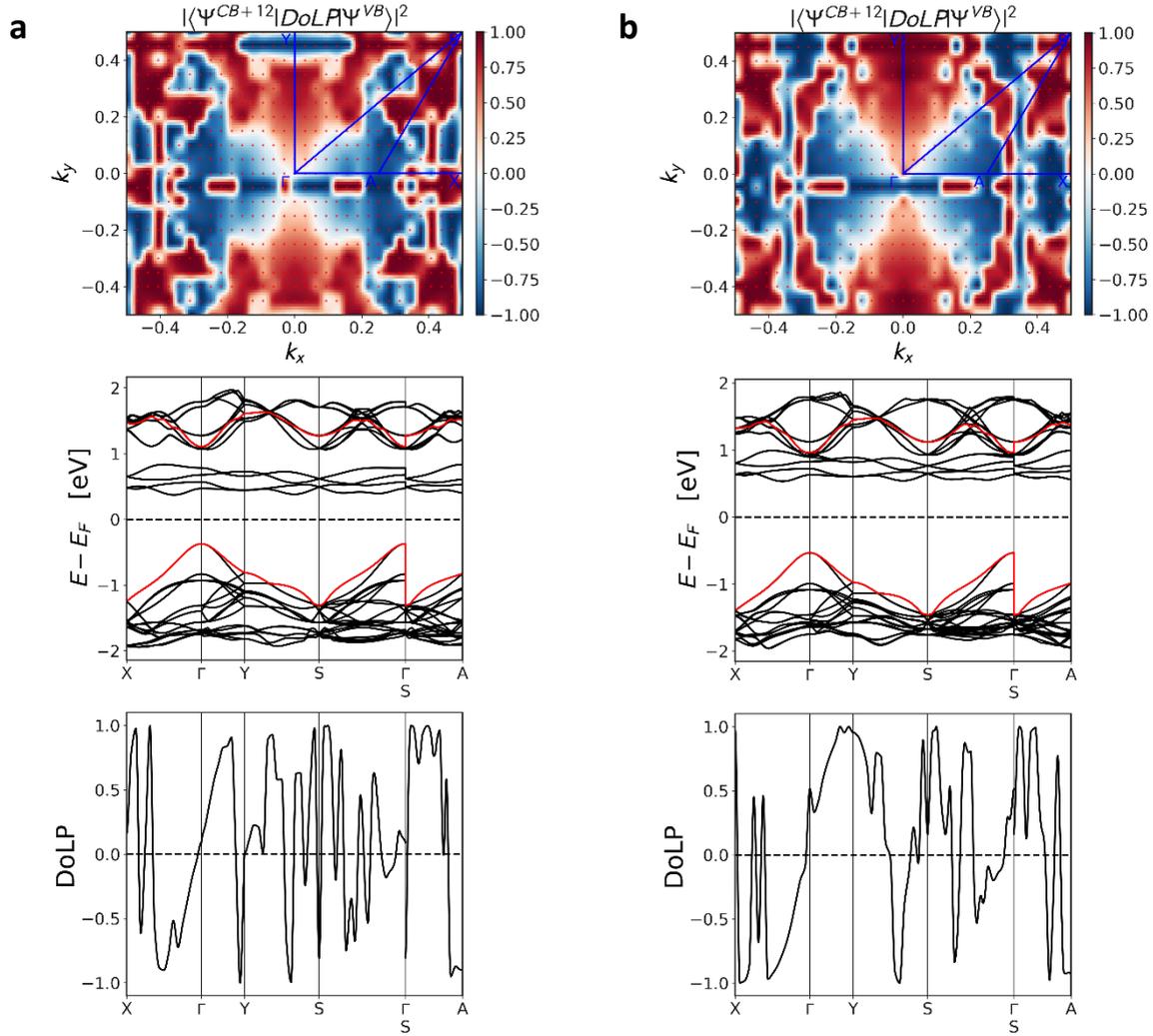

**Figure S15.** Degree of linear polarization (DLP) for the strongest WSe$_2$ intralayer transition (shown in Figure 7a of the main text): **a.** without external electric field, and **b.** with an external electric field, mimicking the internal fields in the samples. A direct comparison reveals that as soon as the NiPS$_3$ bands approach the conduction bands of WSe$_2$, the DLP increase and becomes more asymmetric. In the 2D map of the MMEs, the shape of the red cones in the $\Gamma$-Y direction becomes sharper, and the small red area near $\Gamma$ disappears.

Along the X-$\Gamma$ line, all bands are spin-degenerate, as indicated by the green arrows; therefore, the spin direction is arbitrary (and possible changes are meaningless). However, along the $\Gamma$-Y and S-$\Gamma$ directions, the expectation values of $\langle\sigma_x\rangle$, $\langle\sigma_y\rangle$, and $\langle\sigma_z\rangle$ change upon applying external electric fields of -0.25 eV Å$^{-1}$ and -0.10 eV Å$^{-1}$ for the monolayer WSe$_2$/monolayer NiPS$_3$ and the bilayer WSe$_2$/monolayer NiPS$_3$ system, respectively. This can be seen by the altered color and direction of the arrows. Notably, at and near the $\Gamma$ point, modifications are especially pronounced in the Ni *3d* bands (at lower energies) and in the conduction bands of WSe$_2$.

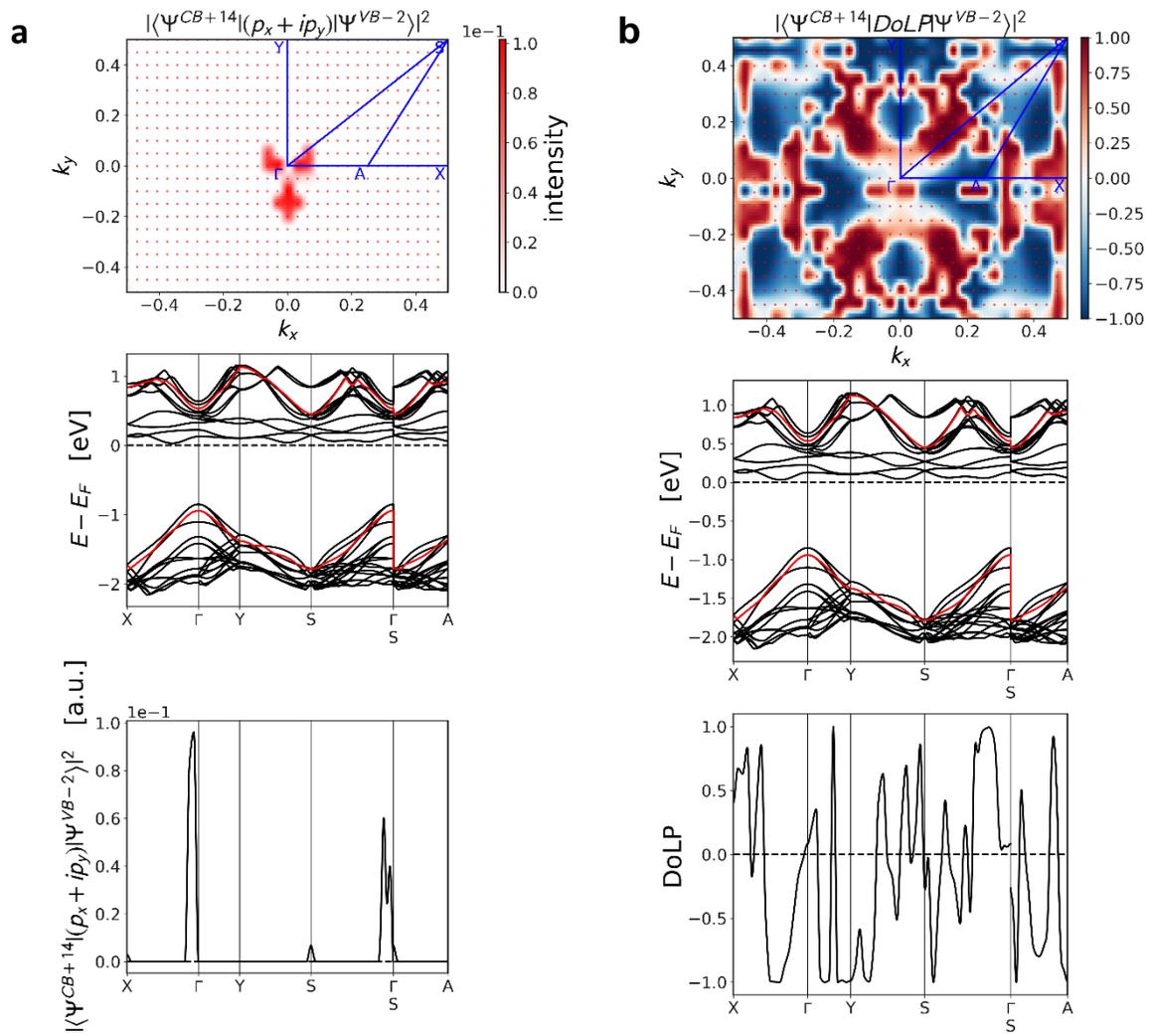

**Figure S16.** One of the strongest WSe$_2$ intralayer transition in the bilayer WSe$_2$/monolayer NiPS$_3$ system. The MMEs for **a**. circularly polarized light, and **b**. DLP. A clear asymmetry in the DLP is observed relative to the $\Gamma$ point, along with a small but finite DLP at $\Gamma$ itself. Since optical excitations do not occur at one specific k point, due to uncertainties such as the finite width of laser pulses or temperature-induced broadening, the transitions, and thus the corresponding excitons display a finite DLP in agreement with the experimental observations.

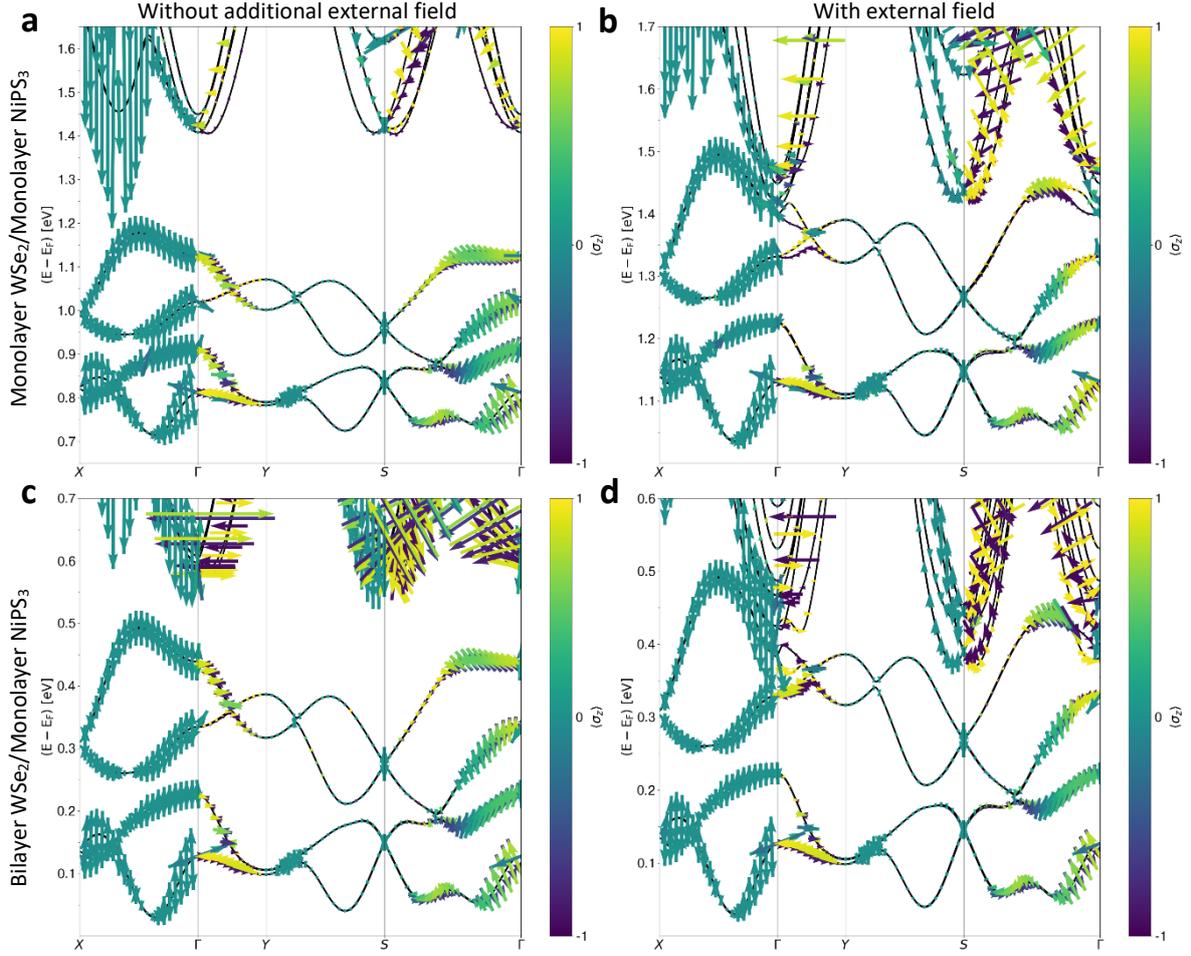

**Figure S17.** Spin texture of the monolayer/monolayer heterostructure in panels **a** and **b**, and the bilayer WSe$_2$/monolayer NiPS$_3$ heterostructure in panels **c** and **d**, shown without (a, c) and with (b, d) an external electric. The expectation value $\langle\sigma_z\rangle$ is indicated by the color cod of the arrows, while the arrow directions indicate the in-plane components, with $\langle\sigma_x\rangle$ and $\langle\sigma_y\rangle$ corresponding to the **k** and energy directions, respectively. Along the X-Γ line, all bands are spin-degenerate, as indicated by the green arrows, and thus the direction of the spins is arbitrary. Yet, along the Γ-Y and S-Γ directions, the expectation values of $\langle\sigma_x\rangle$, $\langle\sigma_y\rangle$, and $\langle\sigma_z\rangle$ change with the application of an external electric field, as can be seen by the changed color and direction of the arrows. Pronounced changes appear at and close the Γ point, particularly in the Ni *3d* bands (at lower energies) and the WSe$_2$ conduction bands.